\documentclass[preprint, amsmath,amssymb, aps,prd,showpacs]{revtex4-1}
\usepackage{epsfig}
\usepackage{dcolumn}% Align table columns on decimal point
\usepackage{bm}% bold math
\usepackage{tikz}
\usepackage{graphicx, graphics, color}
\usepackage{dcolumn}% Align table columns on decimal point
\usepackage{bm}% bold math
\usepackage{hyperref}% add hypertext capabilities
\usepackage[mathlines]{lineno}
\usepackage{float}
\usepackage{subfig}

\newcommand{\be}{\begin{equation}}
\newcommand{\ee}{\end{equation}}
\newcommand{\ben}{\begin{eqnarray}}
\newcommand{\een}{\end{eqnarray}}

\newcommand{\p}{\partial}
\newcommand{\na}{\nabla}

\newcommand{\ep}{\epsilon}

\newcommand{\ga}{\gamma}

\graphicspath{{figs/}}
\graphicspath{{../figs/}}

\keywords{Black Holes}

\begin{document} 

\title{Axion-like dark matter clouds around rotating black holes}

\author{Bartlomiej Kiczek} 
\email{bkiczek@kft.umcs.lublin.pl}
\author{Marek Rogatko} 
\email{rogat@kft.umcs.lublin.pl}
%,marek.rogatko@poczta.umcs.lublin.pl }
\affiliation{Institute of Physics, % \protect \\
Maria Curie-Sklodowska University, % \protect \\
pl.~Marii Curie-Sklodowskiej 1,  20-031 Lublin,  Poland}

\date{\today}% It is always \today, today,

%%%%%%%%%%%%%%%%%%%%%%%
\begin{abstract}
Numerical analysis of {\it dark matter} axion-like cloud in the vicinity of rotating black hole has been performed. The model where 
axion-like scalar field is non-trivially coupled to Maxwell field is studied in the spacetime of Kerr black hole in uniform magnetic field and in Kerr-Newman one. 
The dependence of scalar mass and black hole angular momentum on accumulation of axion {\it dark matter} cloud was given. 
 It was revealed that
condensation of the {\it dark matter} clouds is preferable for a very small mass of axion.

\end{abstract}
%%%%%%%%%%%%%%%%%%%%%%%

\maketitle
\flushbottom

%%%%%%%%%%%%%%%%%%%%%%%%%%%%%%%%%%%%%%%%%%%%%%%%%%%%%%%%%%%%%%%%%%%%%%%%%%%
%%%%%%%%%%%%%%%%%%%%%%%%%%%%%%%%%%%%%%%%%%%%%%%%%%%%%%%%%%%%%%%%%%%%%%%%%%%
\section{Introduction}
\label{sec:intro}
The nature of the elusive ingredient of our Universe {\it dark matter} is one of the most intriguing mysteries of the contemporary physics and astrophysics. 
Ultralight bosons like axion,  axion-like particles,  and {\it dark photons}
could be the answer for these tantalising questions. From the point of view of UV theory, the QCD axions are well motivated as the solution of CP problem \cite{pec77}-\cite{wil78}. 
Recently, axion-like particles widely emerging in the realm of string theory \cite{svr06}, also  attract much attention.

Both axion and axion-like particles are regarded as constituting the possible {\it hidden sector}. This fact triggers the motivation to search them in various kinds of experiments and
theoretical researches.
Namely, it turns out that axion {\it dark matter} has novel effects in polarization of the cosmic microwave background \cite{fed19}
and can be detected in the future terrestrial or astrophysical observations. In Ref. \cite{co19} the new mechanism where a coherently oscillating axion-like 
particle field can transfer its energy to {\it dark photon} has been elucidated. Recently it has been argued that radio telescope observations of neutron stars will enable the
possible detection of axion dark matter, through the axion resonant conversion into radio-frequency photons. The conversion probabilities are proportional to the
strength of magnetic field surrounding neutron star \cite{hua18}-\cite{fos20}.

The process of lasing of an ultralight axion condensate into photons,  relevant for a superradiant axion condensate around stellar mass black hole was elaborated in \cite{sen18}.
The influence of plasma properties placed around black hole in question, on the lasing of axion condensate was also revealed.

It was established \cite{ros18}
that the superradiant instability can lead to the generation of extremely dense axion clouds in the nearby of rotating black holes. Moreover the
stimulated decay may lead to an extremely bright lasers. A possible connection with the observed fast radio bursts was proposed.

Neglecting the rotational effects, axion configuration around pulsars was studied in Ref. \cite{gar18}. Among all it was found that the axions form a localized condensate or radiate
as outgoing waves, depending on the pulsar frequency is smaller or greater that the axion mass.

On the other hand, the analysis of broad-band radio telescope observations of magnetar PSR J1745-2900, enables to establish with the confidence of 95 percent limits, the resonant
axion-photon conversion emission line flux density. These data were translated into limits on axion-photon coupling constant, $g_{a\ga \ga}$ versus axion mass. 
If there is a {\it dark matter} cusp, the limits
reduce to $g_{a\ga \ga} > 6-34 \times10^{-14} GeV^{-1}$, overlapping the axion models with mass range over $33~ eV$ \cite{dar20a}-\cite{dar20b}.
It is argued \cite{ari12}-\cite{gra15}
that the axion coupling to photon depends on the specific model and is related to the values $\sim 10^{-11}-10^{-15} GeV^{-1}$ for intermediate, $\sim 10^{-19} GeV^{-1}$
for Grand Unification Theory and $\sim 10^{-21} GeV^{-1}$ for Planck energy scales.

Studies of light rays passing through an axion and axion-like clouds surrounded  a stationary axisymmetric black hole,
focusing on the experimental setup that is required for the detection of such effect, paying attention to the radio observations of linearly polarized astrophysical sources like active galactic
nuclei, have been performed in \cite{pla18}.

In \cite{gao20} it was proposed to detect axion-like {\it dark matter} by using linearly polarized pulsar light. Pulsar linear polarization angle
may vary with time, due to the birefringence effect which is caused by an oscillating galactic aforementioned {\it hidden sector} component.

The numerical solution of the laser emission problem from axion dense cloud around spinning black hole was presented in \cite{ike19}-\cite{bos19}, where it was 
envisage that the laser emission existed at classical level and the presence of electric charge or rotation leaded to the appearance of the black hole with non-trivial axionic hair.
Moreover the coupling constant of the {\it hidden sector} trigger the strong instabilities affecting superradiant clouds around black hole. On the other hand,
in \cite{car18}
the entire spectrum of the most unstable superradiant modes of a Proca field around Kerr black hole was obtained, as well as, constraints on {\it dark photon} and
axion-like particles were given.

%%%%%%%%%%%%
In our paper we elaborate the subject of the possible existence of axionic {\it dark matter} clouds in the spacetime of stationary axisymmetric black hole.
Numerical simulations based on the axion {\it dark matter} model, where axions are coupled to Maxwell field invariant composed of dual and ordinary $U(1)$-gauge
field strengths, enable us to reveal the basic characteristics of the system in question. We shall pay attention to two cases of black holes, i.e.,
a Kerr black hole in a uniform magnetic field and Kerr-Newman spacetime.

%%%%%
The rest of the paper is organized as follows. In Sec.  II we give a short overview of the axion-like {\it dark matter} portal and provide information about studied black hole backgrounds. 
In the subsections we discuss underlying equations of motion and the problem of free energy for {\it dark matter}
axionic cloud around rotating black holes in question. Sec. III is devoted to the description of the achieved results. Namely we examine the possibilities of condensations of {\it dark matter} in the vicinity of Kerr black hole in a uniform magnetic field and around
stationary axisymmetric Kerr-Newman black hole. In Sec. IV we conclude our investigations. 
Finally, Appendix A contains the relevant technical details concerning the numerical method.

%%%%%%%%%%%%%%%%%%%%%%%%%%%%%%%%%%%%%%%%%%%%%%%%%%%%%%%%%%%%%%%%%%
\section{Axion-like {\it dark matter} sector}
In this section we shall present the basic equations standing behind the axion {\it dark matter} sector model, viewed as the axion-like scalar field coupled to
Maxwell $U(1)$-gauge field. The basic idea lies in the non-trivial axionic coupling to Maxwell strength field invariant constructed from dual and ordinary Maxwell field strengths.
In what follows one investigates the behaviour of axion-like {\it dark matter} clouds surrounded spinning black hole in uniform, say galactic, magnetic field as well as
besieged the Kerr-Newman black hole. For convenience we also refer to them as axions.
To commence with we start with  the Einstein-Maxwell-axion {\it dark matter} theory described by the following action:
\begin{equation}
\mathcal{S} = \int d^4 x \sqrt{-g} \left[R - \frac{1}{4} F_{\mu \nu} F^{\mu \nu}
 - \frac{1}{2} \nabla_\mu \Psi \nabla^\mu \Psi - \frac{\mu^2}{2} \Psi^2
 - \frac{k}{2} \Psi \ast F^{\mu \nu} F_{\mu \nu} \right],
\end{equation}
where $R$ is the Ricci scalar,  $F_{\mu \nu} = 2 \nabla_{[\mu} A_{\nu]}$ is the Maxwell field strength tensor and $\Psi$ is the scalar field  (axion) with mass $\mu$.
The last term of the action describes the coupling of axion field $\Psi$ to one of the electromagnetic field invariants, composed of Maxwell and dual Maxwell filed strengths,
where by $\ast$ we have denoted the Hodge dual operator. $k$ 
constitutes the axionic coupling constant to $U(1)$-gauge field. 

Varying the action with respect to the scalar field $\Psi$ we obtain the equation,
\begin{equation}
\nabla_\mu \nabla^\mu \Psi - \mu^2 \Psi 
- \frac{k}{2} ~\ast F^{\mu \nu} F_{\mu \nu}  = 0.
\label{eq:field_eqn}
\end{equation}
On the other hand, $U(1)$-gauge field is subject to the relation
\be
\na_\mu F^{\nu \mu} + 2 k~\ast F^{\nu \mu} \na_{\mu }\Psi = 0.
\ee
The resulting Klein-Gordon-like equation (\ref{eq:field_eqn})
contains,  despite the standard dynamical and mass terms, an additional source term,  being independent of axion-like field $\Psi$.
The presence of the non-zero source term, containing the dual invariant, explicitly defined as
\begin{equation}
\mathcal{I} = ~\ast F^{\mu \nu} F_{\mu \nu} = \frac{1}{2} \epsilon^{\mu\nu\rho\lambda} F_{\rho \lambda} F_{\mu \nu},
\end{equation}
where $\ep^{\alpha \beta \ga \delta}$ stands for totally antisymmetric Levi-Civita symbol, 
is crucial for the scalarization of a black hole. Namely, if it is equal to zero, the axion-like scalar field equation of motion reduces to the simple massive Klein-Gordon case,
without any self-interaction potential. Then the no-hair theorem plays its role and prevents any scalar hair configuration on the black hole from emerging.

On the other hand, it is easy to check that the invariant in question, $\ast F_{\mu \nu} F^{\mu \nu}$,
is equal to zero in the case when $F_{\mu \nu} =0$, or for spherically symmetric spacetime. In order to be non-trivial, $\ast F_{\mu \nu} F^{\mu \nu} \neq 0$, 
 has to ensure both
rotational and magnetic $U(1)$-gauge field components.

In what follows the main objective of our paper will be to elaborate the behaviour of axion-like field {\it dark matter} sector in the vicinity of a black hole.
As it was remarked the survivability of the $\mathcal{I}$ term in the equation (\ref{eq:field_eqn}) would be crucial for our studies. Therefore we implement
magnetic field in the considered stationary axisymmetric black hole spacetime in two ways. Internally as a consequence of Kerr-Newman black hole solution and externally,
as, e.g., galactic magnetic field surrounding Kerr black hole. The latter idea was originally proposed by Wald in \cite{wal74}, where the uniform magnetic field around black hole was
studied.

We shall consider both of these background line elements and investigate properties of axionic {\it dark matter} clouds around black holes in question.

%%%%%%%%%%%%%%%%%%%%%%%%%%%%%%%%%%%%%%%%%%%%%%%%%%%%%%%%%%%%%%%%%%%%%%%%%%%%%%%%
\subsection{Kerr black hole in  uniform magnetic field} 
\label{sec:kw}
In this section we recall, for the reader's convenience, the basic idea concerning the Wald's introduction of uniform magnetic field in the spacetime of Kerr black hole \cite{wal74}.
The line element of Kerr black hole in Boyer-Lindquist coordinates is provided by the following:
\begin{equation}
ds^2 = -\left(1 - \frac{2 M r}{\Sigma}\right)dt^2 - \frac{4 M r a \sin^2\theta}{\Sigma}dt d\phi + \frac{\Sigma}{\Delta}dr^2 
+ \Sigma d\theta^2 + \frac{\Xi \sin^2\theta}{\Sigma}d\phi^2, 
\label{eq_lineelement}
\end{equation}
with the auxiliary functions defined as
$$\Sigma(r, \theta) = r^2 + a^2 \cos^2 \theta, $$
$$\Delta(r) = r^2 - 2 M r + a^2, $$
$$\Xi(r, \theta) = (r^2 + a^2)^2 -  a^2 \Delta \sin^2 \theta. $$
The solution naturally describes a rotating black hole and is parametrized by two physical quantities, black hole mass $M$ and angular momentum parameter $a = \frac{J}{M}$.
The stationary axisymmetric line element (\ref{eq_lineelement}) possesses two Killing vector fields, the timelike $k_\mu = (\p/\p t)_\mu$
and axial one $m_\mu = (\p/\p \phi)_\mu$.

If we consider the electromagnetic field equations in the spacetime of Kerr black hole,
neglecting the metric back reaction, it is possible to derive a general analytical form of the vector potential,  being a combination of Killing vectors of the underlying spacetime, such as
\begin{equation}
A_\mu = \frac{1}{2} B (m_\mu + 2 a k_\mu ).
\end{equation}
In this way we can introduce a static magnetic field to the system, which is oriented  along the black hole rotation axis.
From astrophysical perspective, such a case may seem quite idealised, however it is an interesting starting point for including magnetic fields into field theories around black holes. 
One way or another, any external (galactic) magnetic field can be cast on the parallel and perpendicular (to the rotation axis) components, and the perpendicular component can be neglected.
Using this set up allows us to utilize all the mathematical properties of the Kerr geometry, such as axial symmetry, in constructing the numerical solution for the scalar.
One has to remember, however, that a non-zero magnetic field breaks the reflection symmetry with respect to equatorial plane.

\begin{figure}[h]
%\centering
\includegraphics[width=0.5\textwidth]{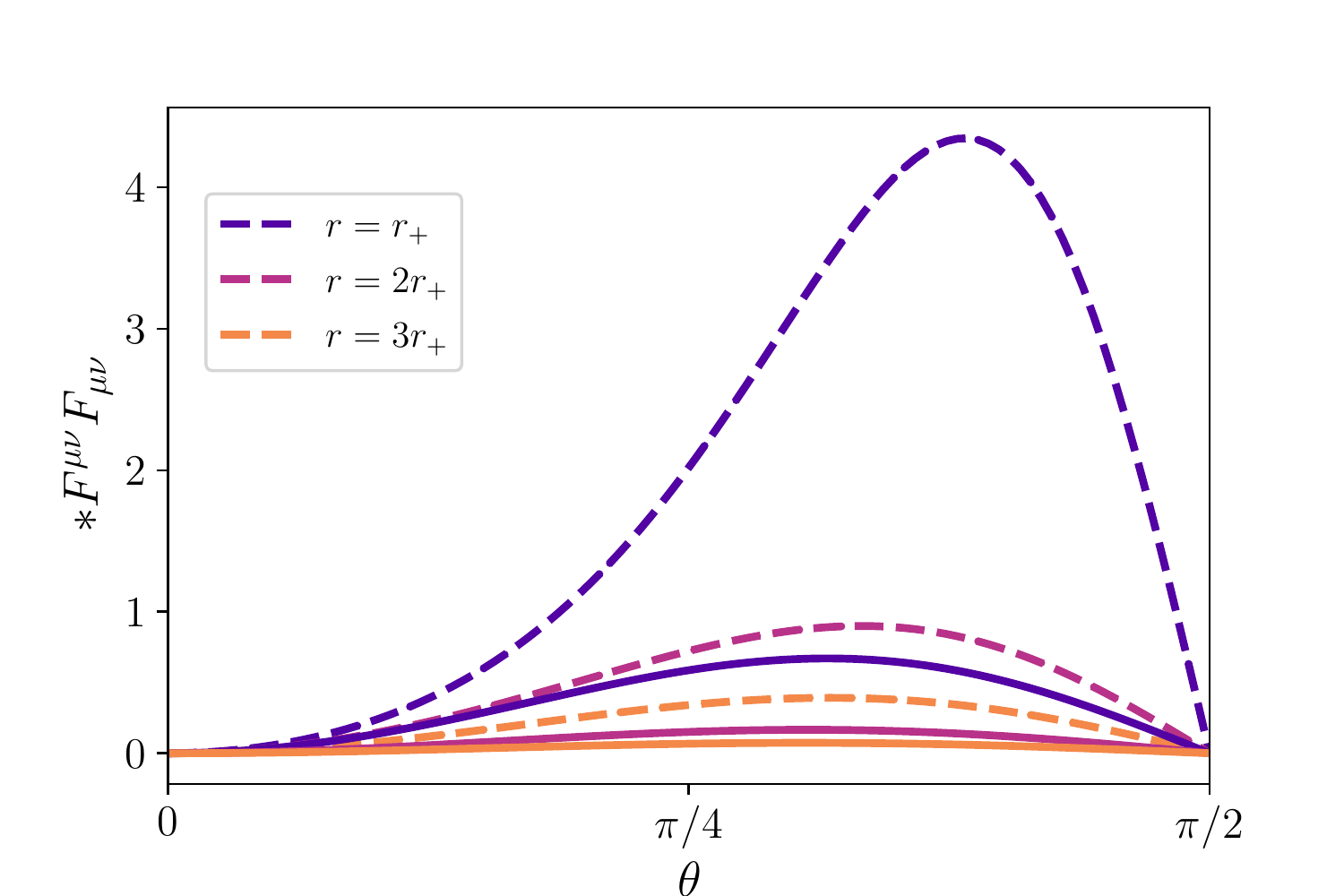}
\caption{Angular and radial dependence of the Maxwell field invariant in Kerr spacetime evaluated at the event horizon.  Colours of the lines indicate subsequent surfaces of constant $r$. The solid lines correspond to $a = 0.5$, while the dashed to $a = 0.99$.  The rise of angular momentum pumps up the value of the invariant significantly and shifts its peak towards black hole equator. }
\label{fig:kw_ff_shape}
\end{figure}

As it has been already mentioned we are interested in static magnetic field, parallel to the rotation axis, so we can drop the timelike Killing vector from the general form of the gauge potential and 
write it in the form as follows:
\begin{equation}
A_\mu dx^\mu = \frac{1}{2} B~ g_{\mu \nu}~ m^{\nu} dx^\mu = \frac{B \sin^2 \theta}{2 \Sigma} \left( -2 M a r dt + \Xi d \phi \right).
\end{equation}
In order to proceed to the analysis of axion {\it dark matter} equation of motion, we should find the invariant $\mathcal{I}$ in the spacetime under consideration.
Its explicit form is following
\begin{align}
\mathcal{I} = - \frac{a B^2 M \sin^2 \theta \cos \theta}{2 \Sigma^4} \big[ 3 a^6 + 2 a^4 M r - 5 a^4 r^2 - 8 a^2 M r^3 - 32 a^2 r^4 - 24 r^6 \nonumber \\ 
+ 4 a^2 (a^4 - a^2 r^2 + 2(M - r)r^3 ) \cos 2\theta + a^4 (a^2 - 2 M r + r^2) \cos 4\theta \big]. 
\end{align}
Because of the fact that the obtained formula is a bit long and complicated and it might not be easy to imagine is shape, for the convenience
of the reader, we visualise it in Fig.  \ref{fig:kw_ff_shape}.

%%%%%%%%%%%%%%%%%%%%%%%%%%%%%%%%%%%%%%%%%%%%%%%%%%%%%%%%%%%%%%%%%%
\subsection{Kerr-Newman black hole spacetime}
As far as the Kerr-Newman spacetime
is concerned, it generalizes Kerr solution and represents a black hole that does not only rotate but also is electrically charged.
The line element implies
\begin{equation}
ds^2 = - \left(1 - \frac{2 M r}{\Sigma} + \frac{Q^2}{\Sigma}\right) dt^2 
- \frac{2 a (2 M r - Q^2) \sin^2 \theta}{\Sigma} dt d\phi
+ \frac{\Sigma}{\Delta} dr^2
+ \Sigma d\theta^2
+ \frac{\Xi \sin^2 \theta}{\Sigma} d\phi^2.
\end{equation}
The auxiliary functions $\Sigma$ and $\Xi$ are defined in the same way as in the previous case, however $\Delta(r) $ has the form given by
\begin{equation}
\Delta(r) = r^2 - 2 M r + a^2 + Q^2,
\end{equation}
where $Q$ is the electric charge of the black hole.
The solution naturally possesses a non-zero electromagnetic vector potential of the form
\begin{equation}
A_\mu dx^\mu = \frac{r Q}{\Sigma}dt - \frac{a r Q \sin^2 \theta}{\Sigma} d\phi.
\end{equation}
On the other hand, 
the corresponding electromagnetic invariant,
needed in the axion {\it dark matter} equations of motion,
 acquires a new, simpler form, described by 
\begin{equation}
\mathcal{I} = - \frac{4 a Q^2 r (a^2 - 2 r^2 + a^2 \cos 2\theta) \cos \theta}{\Sigma^4}.
\end{equation}

%%%%%%%%%%%%%%%%%%%%%%%%%%%%%%%%%%%%%%%%%%%%%%%%%%%%%%%%%%%%%
%%%%%%%%%%%%%%%%%%%%%%%%%%%%%%%%%%%%%%%%%%%%%%%%%%%%%%%%%%%%%
\subsection{Equation of motion}
Let us suppose that the axion {\it dark matter} field will be a function depending on two coordinates, radial and azimuthal ones, i.e.,
$
\Psi = \psi(r, \theta).
$
It leads to the axion equation of motion provided by the relation
\begin{equation}
\Delta \partial_r^2 \psi + 2(r - M) \partial_r \psi + \partial^2_\theta \psi + \cot \theta~ \partial_\theta \psi - \mu^2 \Sigma \psi = \frac{k \Sigma}{2} \mathcal{I}, \label{axion_eqn}
\end{equation}
$\p_m$ stands for the derivative with respect to $m$-coordinate.
The obtained equation is an elliptic partial differential equation, which is linear in $\psi$.
The general form of this equation remains the same in both backgrounds. They differ by the shape of $\Delta$ function, which can be enriched with the electric charge,  and by the source term on the right hand side.
It turns out that 
the equation \eqref{axion_eqn} follows a scaling transformation of the form
\begin{align*}
r ~\rightarrow cr \qquad a \rightarrow c a \\
M ~\rightarrow cM \qquad k \rightarrow k c^2 \\
Q ~\rightarrow c Q \qquad B \rightarrow B / c \\ 
\mu^2 \rightarrow \mu^2/c^2,
\end{align*}
which allows us to fix one quantity to unity.
The scaling concerns the quantities from both backgrounds.
For the convenience, in our numerical simulations, we use the above scaling and fix the black hole mass to unity, $M = 1$.

%%%%%%%%%%%%%%%%%%%%%%%%%%%%%%%%%%%%%%%%%%%%%%%%%%%%%%%%%%%%%%%%%%%%%%%%%%%%%%%%%%%%%
\subsection{Free energy of {\it dark matter} axionic cloud in the spacetime of rotating black hole} 

The existence of the solution to the field equation, i.e.,  some state of the system, does not guarantee that it is the physically preferable configuration.
To verify this, one ought to consider the thermodynamics of the system and look for the relevant quantities \citep{kic20}.
As we consider the gravitational system without backreaction,  the thermodynamical quantities of the black hole such as entropy and Hawking temperature remain unaffected by 
the axionic {\it dark matter} condensate.
Thus we wish to examine the free energy difference generated by the non-trivial profile of the scalar $\psi$ with respect to hairless solution.
To proceed further,  we
consider the $\psi$ dependent part of the underlying action,
\begin{equation}
\mathcal{S}_{axion} = \int d^4 x \sqrt{-g} \left[- \frac{1}{2} \nabla_\mu \Psi \nabla^\mu \Psi - \frac{\mu^2}{2} \Psi^2
 - \frac{k}{2} \Psi  \ast F^{\mu \nu} F_{\mu \nu} \right].
\label{eq:axion_only_action}
\end{equation}
In order to find the free energy contribution of the axionic cloud, we evaluate the Euclidean on-shell action related with \eqref{eq:axion_only_action}.

Firstly, we use the equation of motion for the axionic field $\psi$ \eqref{eq:field_eqn} and substitute it into the action. This allows us to remove the term with the Maxwell field strength tensor.
After few transformations we make use of Gauss theorem and split the action into volume and surface terms.
Because of the boundary conditions (see the next section for details) the surface integral vanishes.
In last step we perform Wick's rotation of the time coordinate and get the explicit formula provided by
\begin{equation}
F = - 2 \pi \int_\mathcal{M} dr d\theta ~\frac{\Sigma \sin \theta}{2} \bigg[ (\partial_r \psi)^2 g^{rr} + (\partial_\theta \psi)^2 g^{\theta \theta} + \mu^2 \psi^2 \bigg].
\label{eq_freeenergy}
\end{equation}

As the integrand is positive, in the whole domain, the free energy shift is negative for every configuration of $\psi$ field being the solution to the considered equations of motion.
Although there is a caveat.
It can be supposed that any non-trivial $\psi$ will be preferred by nature.
However this is not really the case.  For the given ansatz the system has only trivial zero solution when the source term $\mathcal{I}$ is zero.
This allows us to state that the considered axionic {\it dark matter} clouds are magnetically induced, and are only present in the system when $\mathcal{I}$ is non-trivial.

The equation \eqref{eq_freeenergy} will be extensively exploited in the following section,  to achieve the free energy plots. The aforementioned integral will be computed numerically.

%%%%%%%%%%%%%%%%%%%%%%%%%%%%%%%%%%%
%%%%%%%%%%%%%%%%%%%%%%%%%%%%%%%%%%%%%%%%%%%%%%%%%%%%%%%%%%%%%%%%%%%%%%%%%%%
\section{Numerical results}
This section will be devoted to the obtained numerical solutions of both axion {\it dark matter} clouds surrounding
Kerr black hole immersed in a uniform magnetic field and Kerr-Newman one.
As we have already mentioned they differ by the shape of the source term originating from axionic coupling and the metric function $\Delta$.
We deal with the partial differential equation \eqref{axion_eqn} by virtue of Chebyshev spectral methods.
%First we discretize it, on Chebyshev grid and solve the obtained matrix system with a help of the standard linear algebra methods.  
As the equation is fully linear in $\psi$, the acquired solution is unique and well defined.
For technical details of the numerical method see Appendix A.
%The error of calculated solution is estimated to be $< 10^{-5}$.

In the above set up one considers the field only above the black hole event horizon (including the horizon itself). The bounds for 
the radial coordinate are from the event horizon to spatial infinity, precisely $r \in [r_+, \infty)$.
The symmetry of the spacetime allows us to narrow the domain to one quarter of the $(r, \theta)$ plane.  For convenience we pick the first quarter, with $\theta \in [0, \pi/2]$.
Values of the solution for remaining quarters can be achieved by the negative reflection with respect to the equatorial plane 
\begin{equation}
\psi(r, \theta - \pi/2) = - \psi(r,  \pi/2 - \theta),
\end{equation}
and the remaining part of the solution can be obtained by the rotation.
Having our numerical domain defined we can move to the necessary transformations.

To implement spectral methods based on Chebyshev polynomials we have to map the coordinates of the manifold onto $[-1, 1]$ intervals. 
In order to do this we use following transformations for $r$ and $\theta$
\begin{equation}
z = 1 - \frac{2 r_+}{r},
\end{equation}
\begin{equation}
u =  \frac{4\theta}{\pi} - 1,
\end{equation}
where $r_+ = M + \sqrt{M^2 - a^2 - Q^2}$, which is the standard definition of the outer event horizon of rotating black holes.  After
the transformation $z = -1$ represents the inner boundary - the event horizon and $z = 1$ spatial infinity.
Similarly, for $u = -1$ one thinks about a north pole of a black hole, while $u = 1$ represents the equator of the object in question.

%%%%%%%%%
Let us now discuss the boundary condition imposed on the solution of the axionic {\it dark matter} field equation.
Namely, 
for the black hole event horizon $r = r_+$ we demand that the derivative with respect to
 $r$-coordinate is given by 
 $\partial_r \psi = 0$. This fact ensures the regularity of the solution. For $r \rightarrow \infty$, the field equation takes the simple, angle independent form provided by the relation
\begin{equation}
\partial^2_r \psi + \frac{2}{r} \partial_r \psi - \mu^2 \psi = 0,
\end{equation}
which implies that
\begin{equation}
\psi(r) \sim \frac{e^{-\mu r}}{r} + \frac{e^{\mu r}}{r}.
\label{psi2}
\end{equation}   
 The asymptotic flatness and regularity cause that the second term in the relation (\ref{psi2}) vanishes.  For a finite mass solution vanishes in the infinity.  Thus, the requirement that
  $\psi(r \rightarrow \infty) = 0$ comprises the second boundary condition.

In case of the boundaries imposed on the polar angle $\theta$, we use the argumentation based on the symmetries of the spacetime. 
For the north pole $\theta = 0$, the axial symmetry of the rotating black hole obliges the solution to be invariant under the transformation $\phi \rightarrow \phi \pm \pi$.
{In other words the solution ought to be even along a meridian, with respect to the pole.}
Therefore $\partial_\theta \psi = 0$ is the reasonable choice.
% Equator
However for the equatorial plane $\theta = \pi/2$ the situation is different, as it constitutes the place where both source terms change signs, and so does the solution.
For that reason we demand that $\psi = 0$, there.

%%%%%%%%%%%%%%%%%%%%%%%%%%%%%%%%%%%%%%%%%%%%%%
\subsection{Axionic {\it dark matter} clouds around Kerr black hole in uniform magnetic field}
Now we proceed with the conclusions achieved from the analysis of the numerical solutions of the equation \eqref{axion_eqn} in the adequate
spacetimes of the rotating black holes. To commence with one considers the results obtained for Kerr black hole in uniform magnetic field.

\begin{figure}[h]
\centering
\subfloat[$\mu^2 \rightarrow 0^+, a = 0.5$]{
\includegraphics[width=0.45 \textwidth]{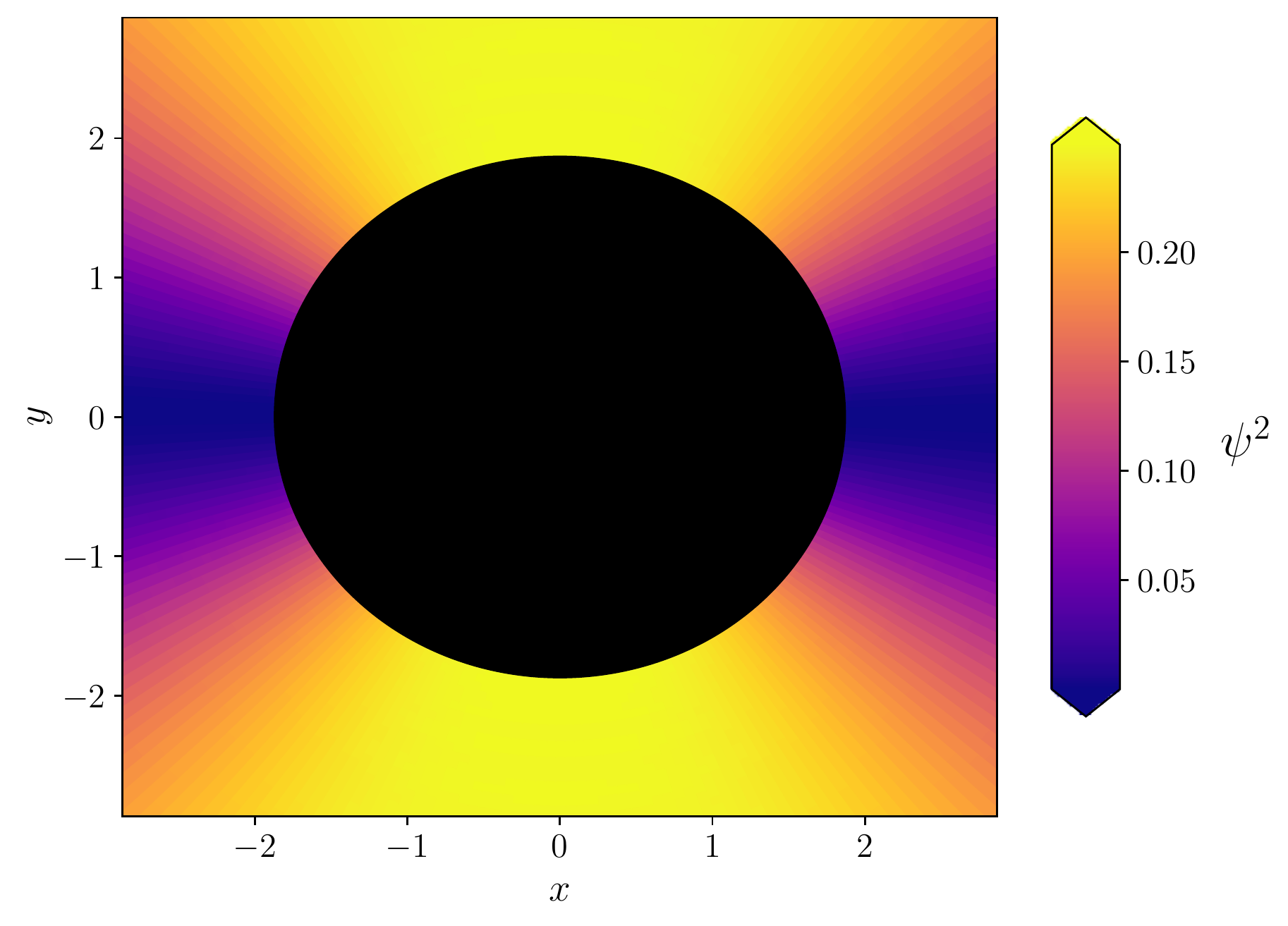}
}
\qquad
\subfloat[$\mu^2 \rightarrow 0^+, a = 0.996$]{
\includegraphics[width=0.45 \textwidth]{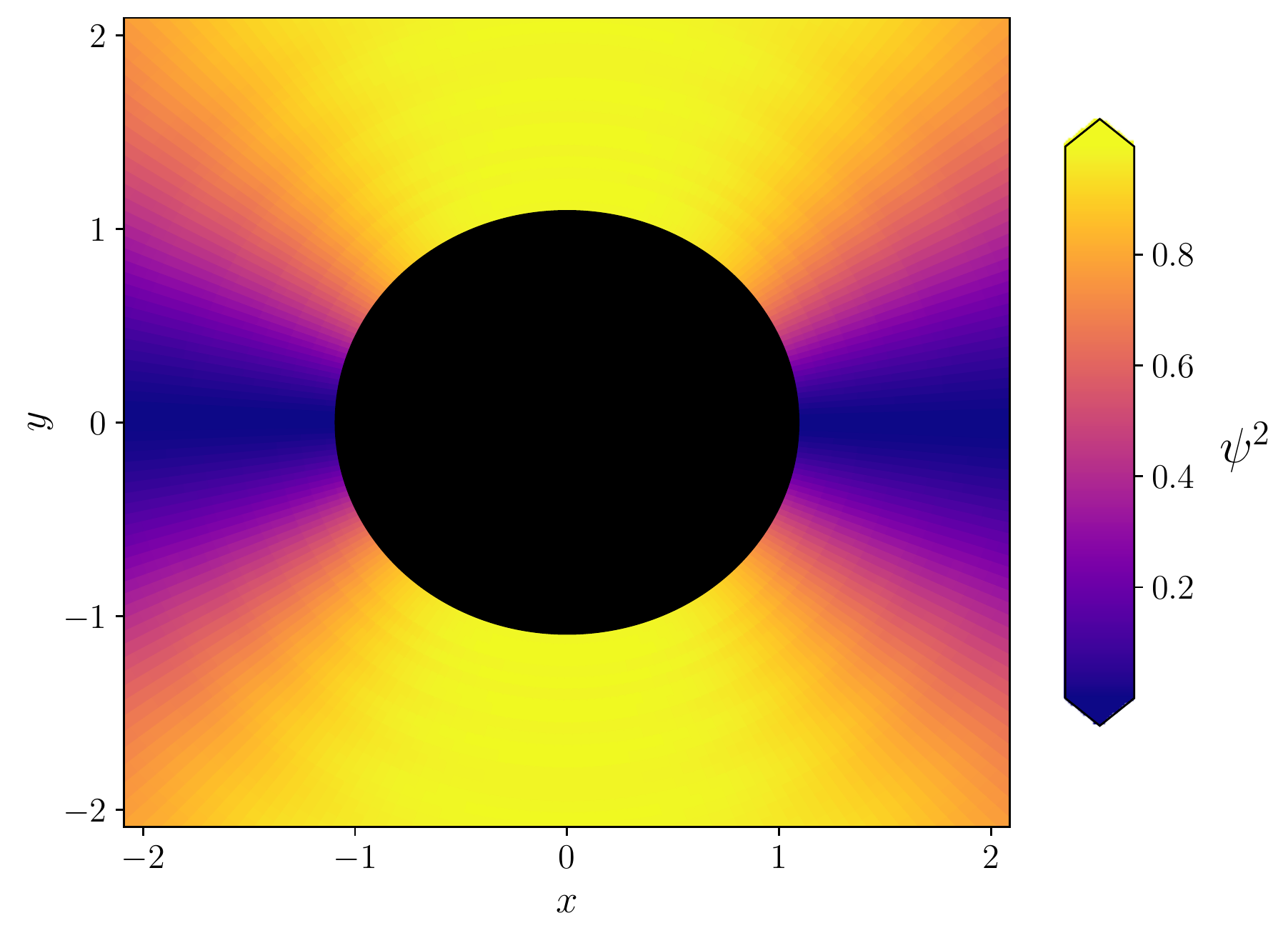}
}

\vspace{0.5cm}
\subfloat[$\mu^2 = 16, a = 0.5$]{
\includegraphics[width=0.45 \textwidth]{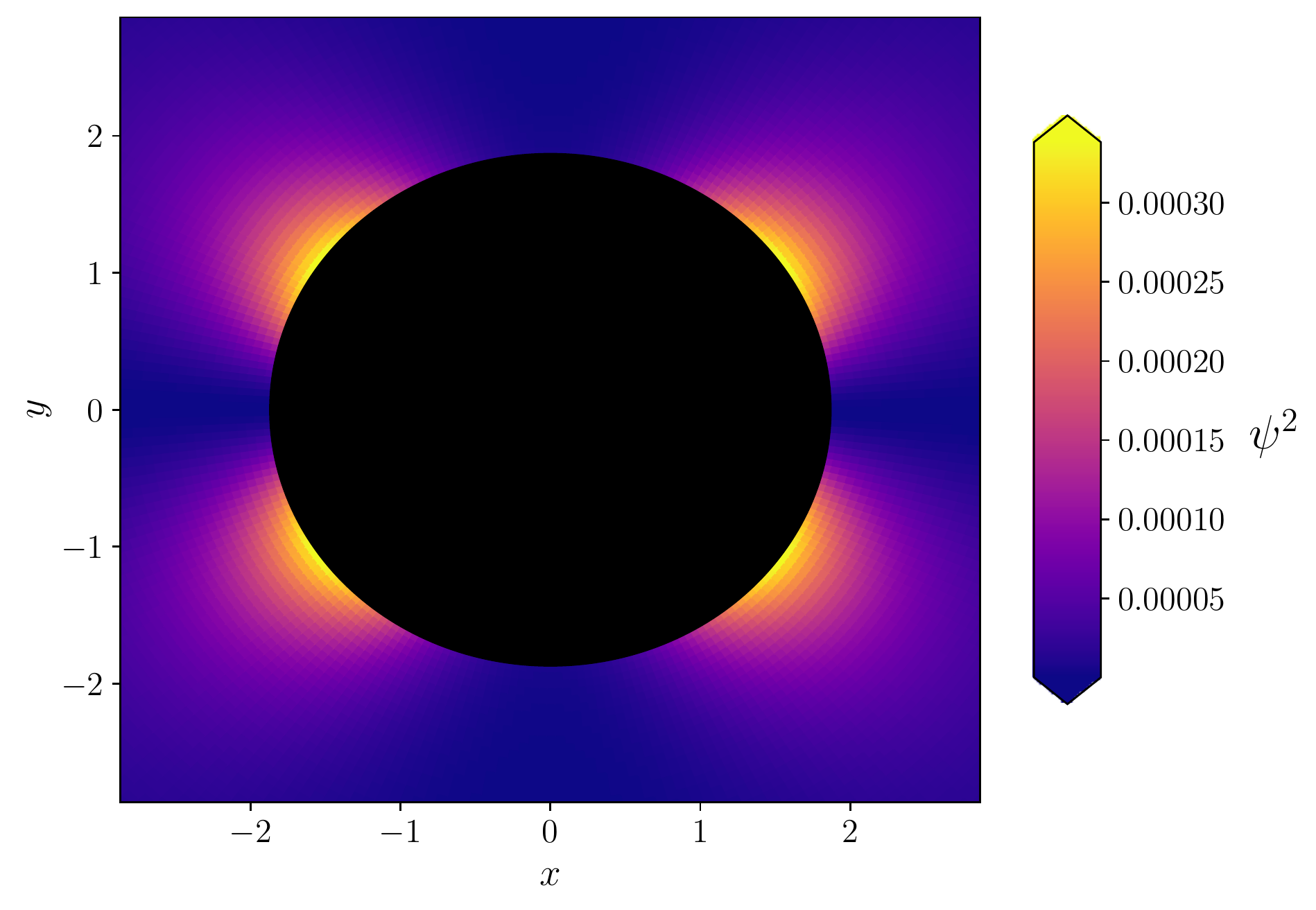}
}
\qquad
\subfloat[$\mu^2 = 16, a = 0.996$]{
\includegraphics[width=0.45 \textwidth]{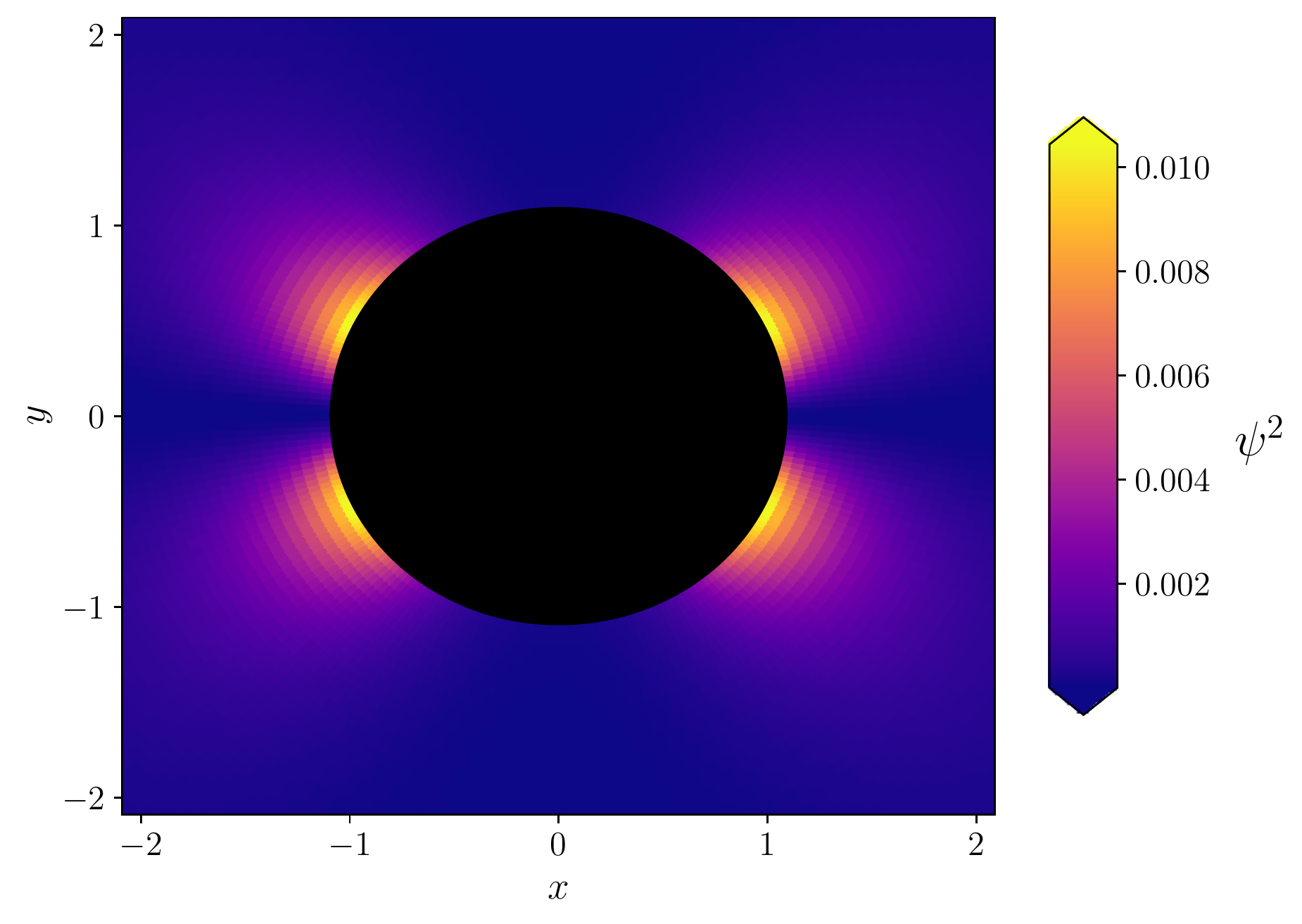}
}
\caption{The distribution of the axionic {\it dark matter} clouds around a black hole with external uniform magnetic field. Every panel corresponds to different values of 
field mass and black hole angular momentum.
For ultra-light mass the cloud concentrates in polar areas,  while the large mass field is dragged towards the equator.}
\label{fig_kw_maps}
\end{figure}

In Fig. \ref{fig_kw_maps} we present first series of spatial distribution plots of axionic {\it dark matter} field.
For the convenience we present the squared distribution $\psi^2$, where the $(r, ~\theta)$ plane has been cast into Cartesian coordinates $(x,~y)$.
The black circle corresponds to the area hidden under the black hole's event horizon.
In the first pair of plots, depicted in panels (a) and (b), one can see the distribution of the ultra-light axionic {\it dark matter} field. 
It can be noticed that it is loosely concentrated around polar regions.
The equatorial plane remains free from the scalar $\psi$, which results from the imposed boundary conditions and the symmetry of the problem in question.
The increase of the black hole angular momentum does not drastically change the field distribution.
However, it significantly influences its magnitude (see the values on colorbars).
Furthermore, the field slowly decays with distance and spreads up to $r = 10r_+$ outside the event horizon. 
This is not the case for the massive axionic {\it dark matter} field, which is presented in panels (c) and (d) of Fig. \ref{fig_kw_maps}. 
The mass $\mu^2$ is quite large, in terms of geometrical units, however it serves the purpose of visualising the contrast.
In the case under inspection the increase of the angular momentum causes dragging the axionic clouds towards the equatorial plane.
This effect seems to be quite intuitive,  by the analogy to the centrifugal forces in classical mechanics.
Similarly to the former case, the angular momentum increase causes a boost in the field $\psi$ magnitude, by several orders.

It should be noted that approximated analytical solution for this background has been derived in Ref. \cite{bos19}.
Our numerical solution perfectly matches the results obtained before, in the considered limit -- a slow rotation of the black hole and zero mass of the axion field.

\begin{figure}[h]
\centering
\includegraphics[width=0.5\textwidth]{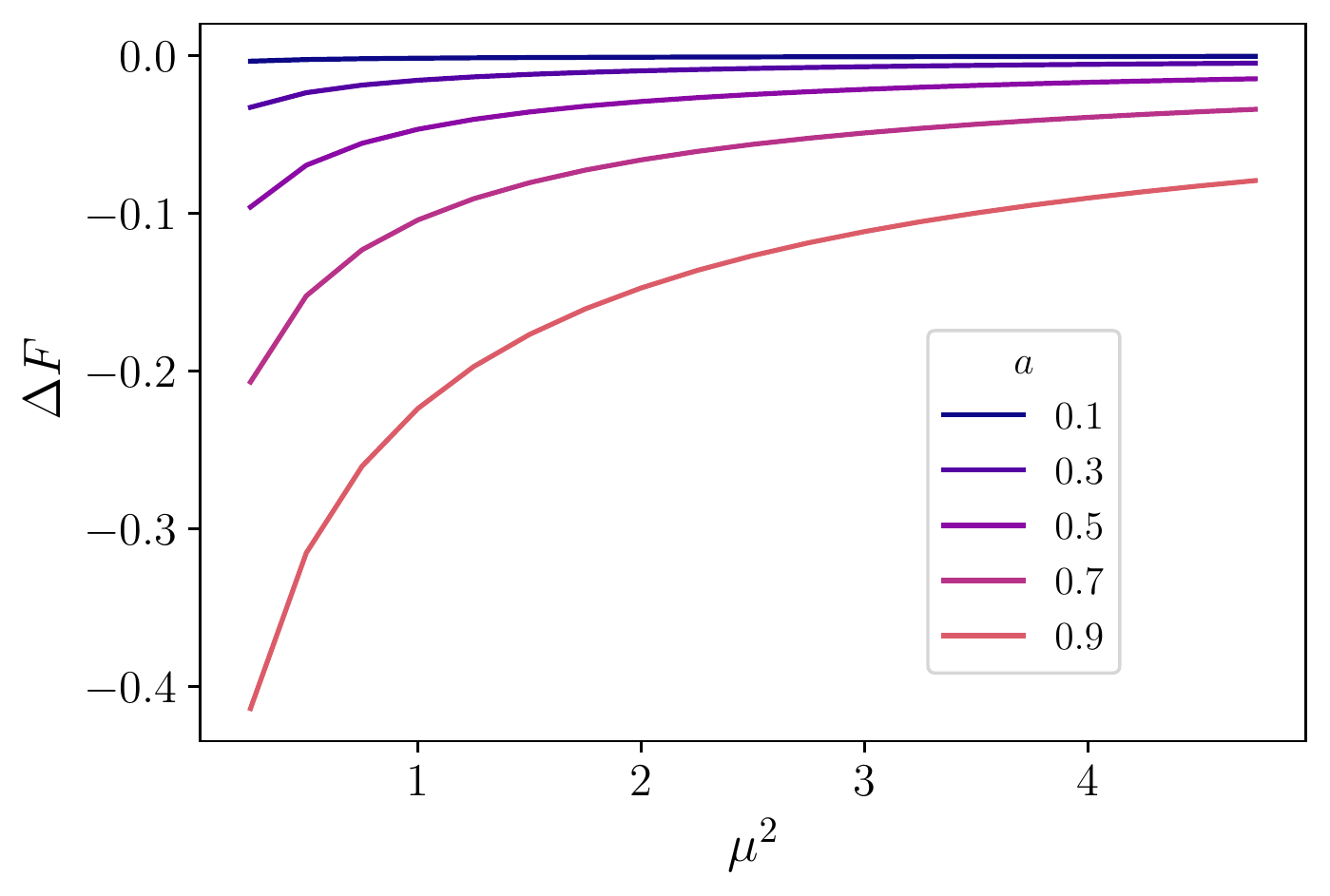}
\caption{Free energy shift as a function of axion-like {\it dark matter} mass. Ultra-light particles are the most preferred ones as they cause the most significant free energy decrease.}
\label{fig_kw_fm}
\end{figure}

To proceed further with the studies of hairy configurations in Kerr-Wald spacetime, let us take a look at the behaviour of the free energy, with respect to the change of the other parameters 
of the theory.
In Fig.  \ref{fig_kw_fm} the free energy shift as a function of the axion field mass can  be observed. Here we present several curves for different values of the black hole angular momentum.
All curves share the similar behaviour, which is scaled differently.
Moreover what all curves have in common is a significant decrease of free energy for small values of the axion mass. 
It means that axionic {\it dark matter} clouds are the most stable and the most strongly bound for very small, almost zero,  field masses.

One might notice an interesting correlation of this result with theoretical predictions for {\it dark matter} axion mass, to be ultra-light in sub eV region. 
Namely, the recent constraints on bosonic {\it dark matter} for ultra-low field nuclear magnetic resonance were proposed \cite{gar19}. The new experimental bounds
for axion-like {\it dark matter} particles, are ranging from $1.8 \times 10^{-16}$ to $7.8 \times 10^{-14}$ eV. Recently the direct implications on the mass of ultra-light {\it dark matter} particles 
by studies of mass and spin of accreting and jetted astrophysical black hole have been established \cite{una20}. It was revealed that axion-like particles with the mass range
$10^{-21} < \mu(eV) < 10^{-19}$ could contribute at almost 10 percent of the {\it dark matter} mass. On the contrary, for the mass range $10^{-19} > \mu(eV) > 10^{-17}$, they
constitute only $0.01$ to $1$ percent of the {\it dark sector} mass.

\begin{figure}[h]
\centering
\includegraphics[width=0.5\textwidth]{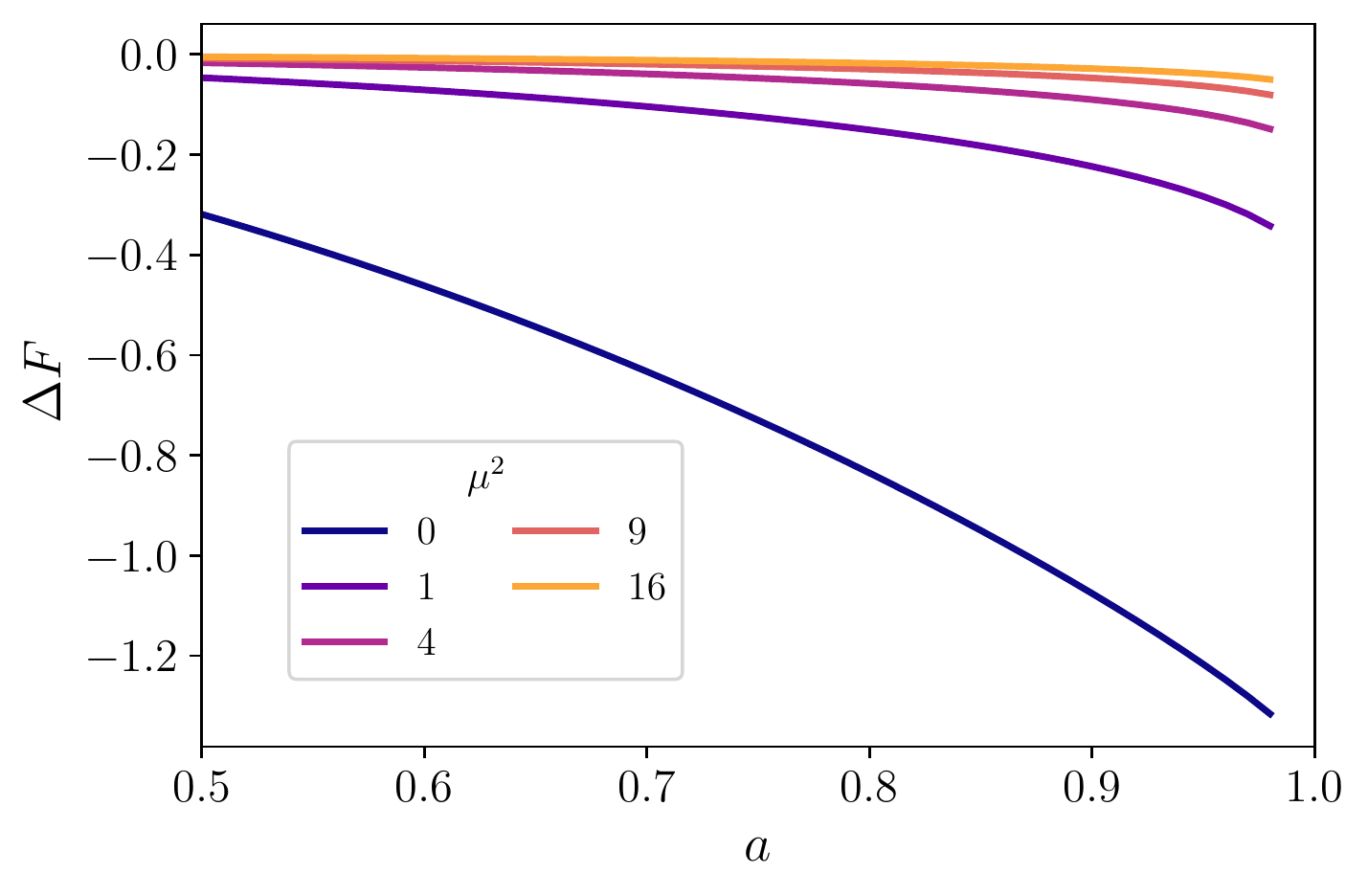}
\caption{Free energy shift vs. the black hole angular momentum. The extremal black holes (with very high angular momentum) constitute a perfect environment for {\it dark matter} axionic 
hair since the free energy fall off is the biggest.}
\label{fig_kw_fa}
\end{figure}

On the other hand Fig. \ref{fig_kw_fa} presents the free energy shift as a function of  black hole angular momentum for different masses of the field $\psi$.
The course of the curves is also very similar and surely they follow some kind of $\mu^2$ dependent scaling. The shift is slight for moderate values of the angular momentum 
and becomes stronger for quickly rotating black holes.
Extremal black holes, with $a$ approaching to $1$, bring the biggest fall off of free energy.
Once again the decrease is the most drastic for the ultra-light field (dark blue curve in the plot).
All these observations allow us to conclude that extremal black holes constitute good environments for the emergence of the ultra-light axionic {\it dark matter} clouds.

%%%%%%%%%%%%%%%%%%%%%%%%%%%%%%%%%%%%%%%%%%%%%%%%%%%
\subsection{Axion {\it dark matter} clouds in the vicinity of Kerr-Newman black hole}

\begin{figure}[h]
\centering
\subfloat[$\mu^2 \rightarrow 0^+, a = 0.5$]{
\includegraphics[width=0.45 \textwidth]{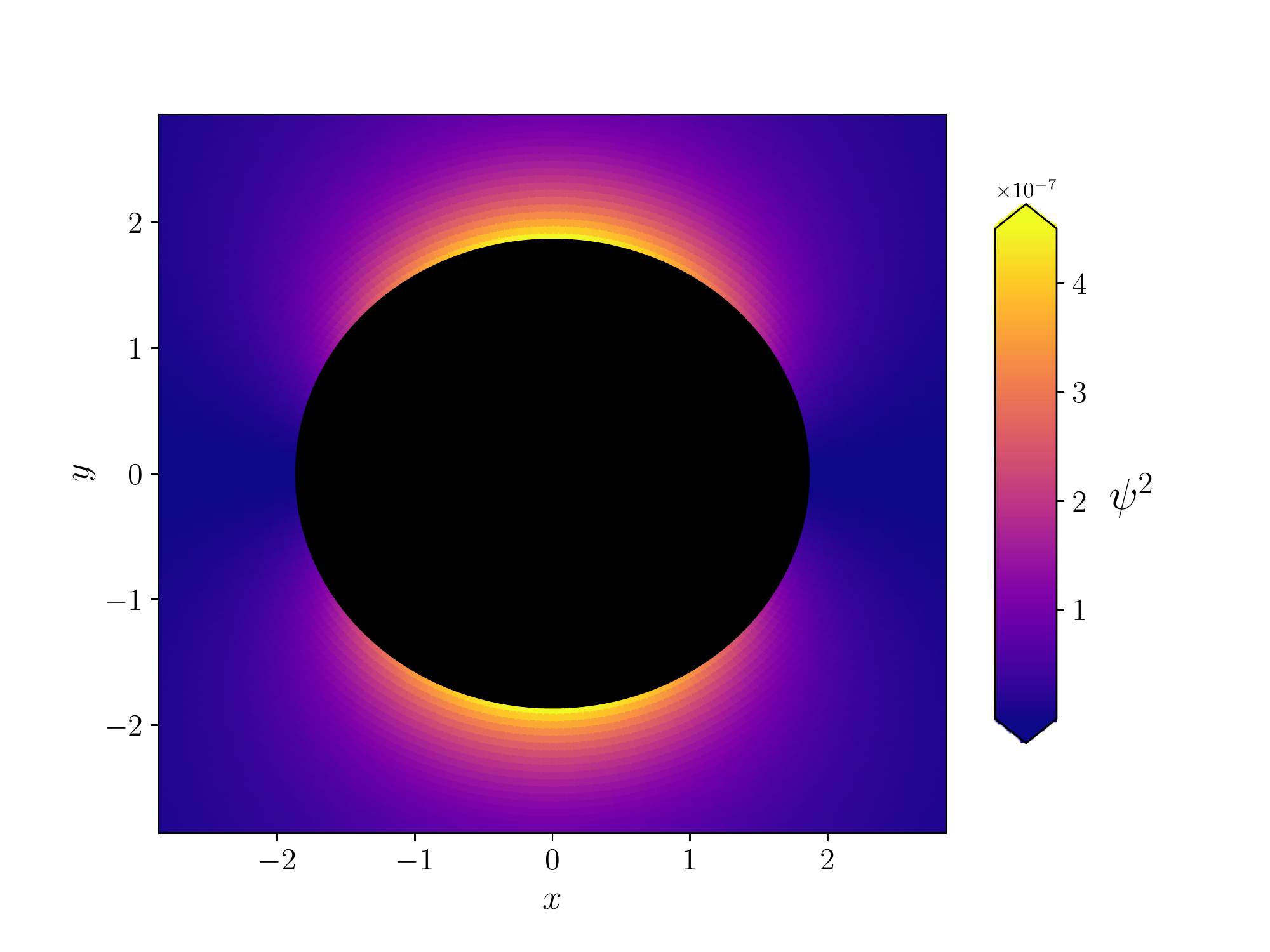}
}
\qquad
\subfloat[$\mu^2 \rightarrow 0^+, a = 0.99$]{
\includegraphics[width=0.45 \textwidth]{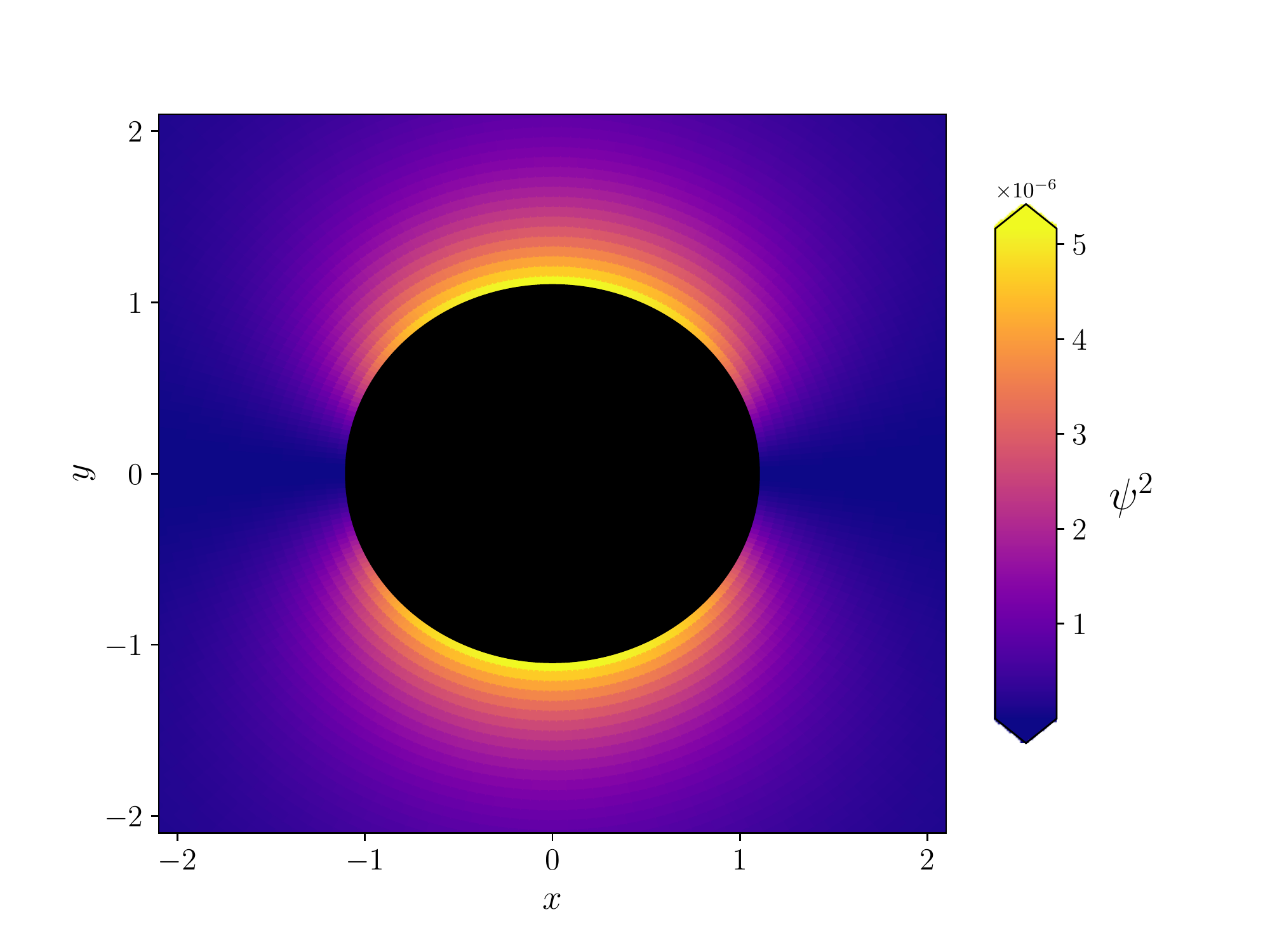}
}

\vspace{0.5cm}
\subfloat[$\mu^2 = 16, a = 0.5$]{
\includegraphics[width=0.45 \textwidth]{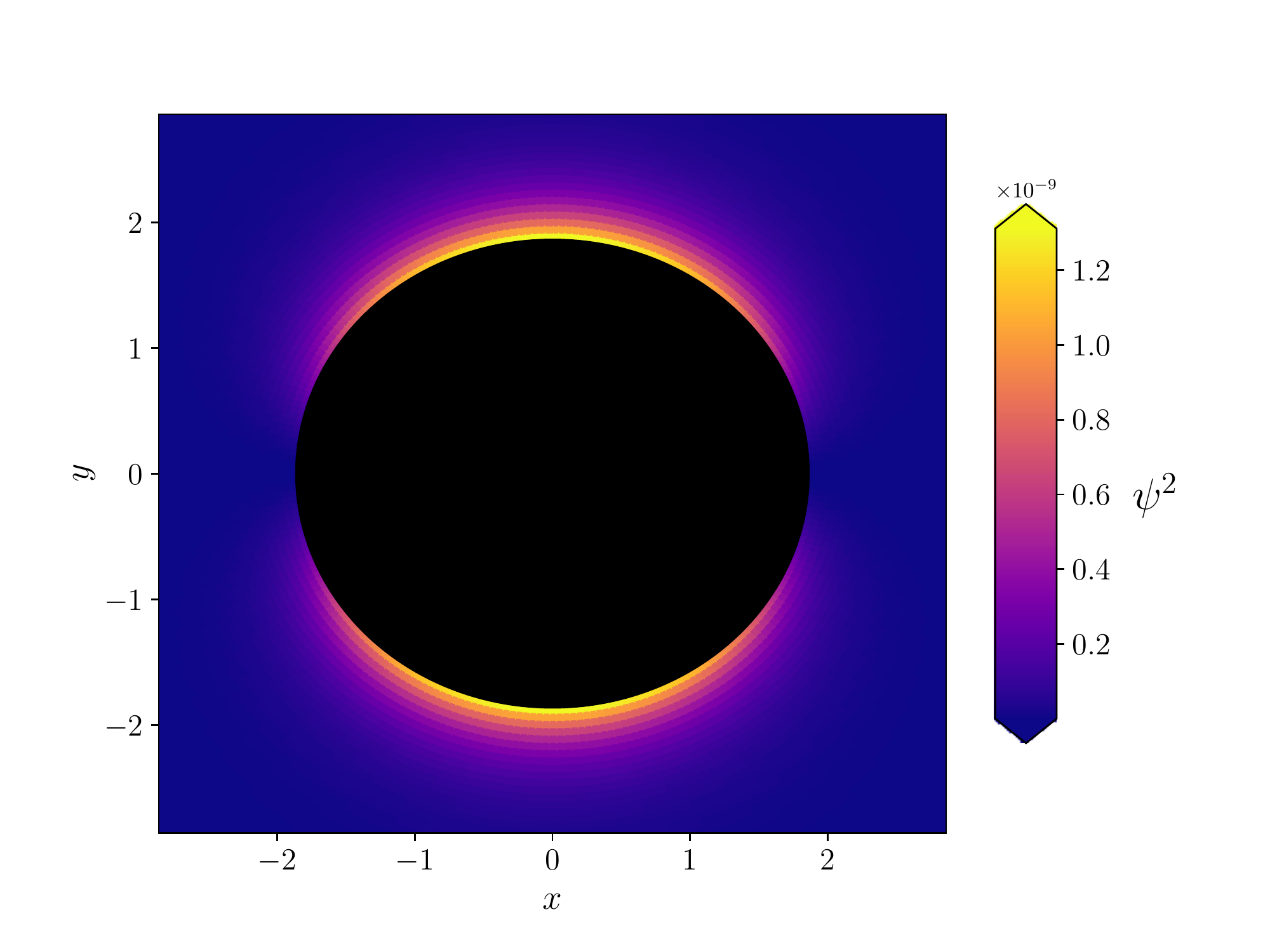}
}
\qquad
\subfloat[$\mu^2 = 16, a = 0.99$]{
\includegraphics[width=0.45 \textwidth]{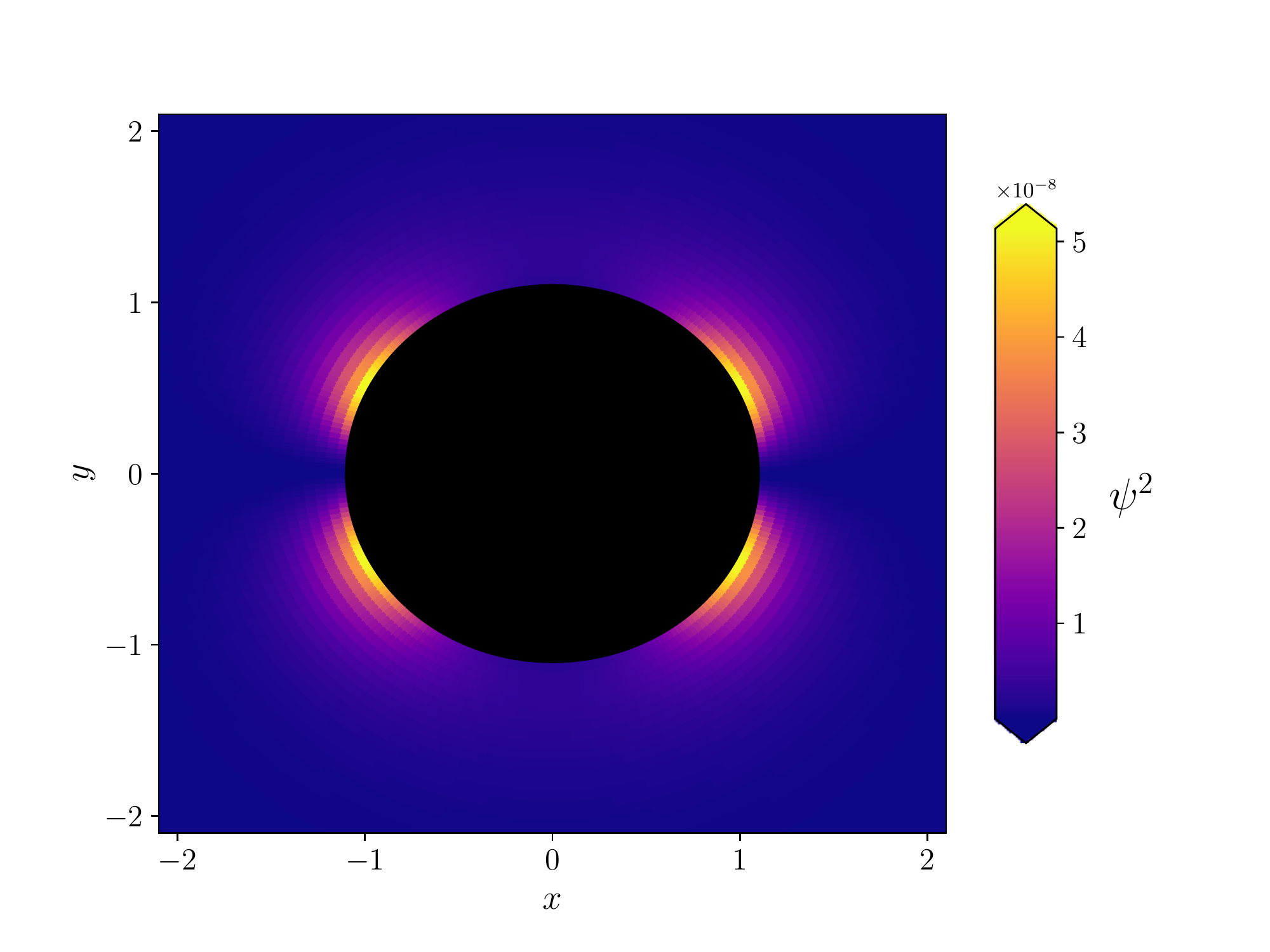}
}
\caption{Axionic {\it dark matter} clouds around Kerr-Newman black holes with $Q = 0.1$.  High angular momentum of the black hole reveals rich geometrical structure of the clouds.}
\label{fig_kn_maps}
\end{figure}

Let us now discuss the characteristic features of axionic {\it dark matter}
clouds nearby Kerr-Newman black holes.
This background has a distinct electromagnetic vector potential, thus the behaviour of axionic hair differs significantly from the former case.
It can be seen in Fig. \ref{fig_kn_maps}, where we plot analogical spatial distributions of $\psi^2$, just like in the case of
Kerr black hole dipped in a uniform magnetic 
background.
However, these solutions are essentially various from two main reasons.
Firstly, the shape of $\mathcal{I}$ expression is different, hence the source term envisages the other kind of solution.
Secondly, the electromagnetic component (the electrical charge $Q$), enters the geometric relations, such as the radius of the event horizon.
Because of this interplay between $a$ and $Q$, the black hole angular momentum is limited to the value below one,  because of the fact that we require $r_+$ to be a real number.
In this illustrative example one sets $Q = 0.1$.
The panels (a) and (b) illustrate ultra-light axionic {\it dark matter} clouds. In the case under inspection,
their distribution is slightly affected by the black hole angular momentum.
When its value is moderate, the cloud aggregates around polar regions of Kerr-Newman black hole.  As the angular momentum rises,  axions flow towards the equator and are spread over whole hemispheres.
The clouds of {\it dark matter} are distributed on the majority of slice's area, except the equatorial region, which is naturally the result of the imposed boundary conditions.
However, contrary to previous background, axionic clouds are strongly localised in the vicinity of black hole event horizon. Their distribution quickly vanishes with the growth of the distance.
The large mass case, presented on panels (c) and (d), reveals a strong concentration of the field for both values of the angular momentum.
In the former case the cloud is visible over the majority of black hole hemispheres while in the latter case it is mostly present in the equatorial area.
Moreover,  the axion {\it dark matter} field blurs further into the space.
The rise of mass also shrinks the spread of the clouds, as they quickly decay with distance from the horizon.  Besides the rise of angular momentum significantly increases the magnitude of the axionic field (see colorbars).

\begin{figure}[h]
\centering
\includegraphics[width=0.5 \textwidth]{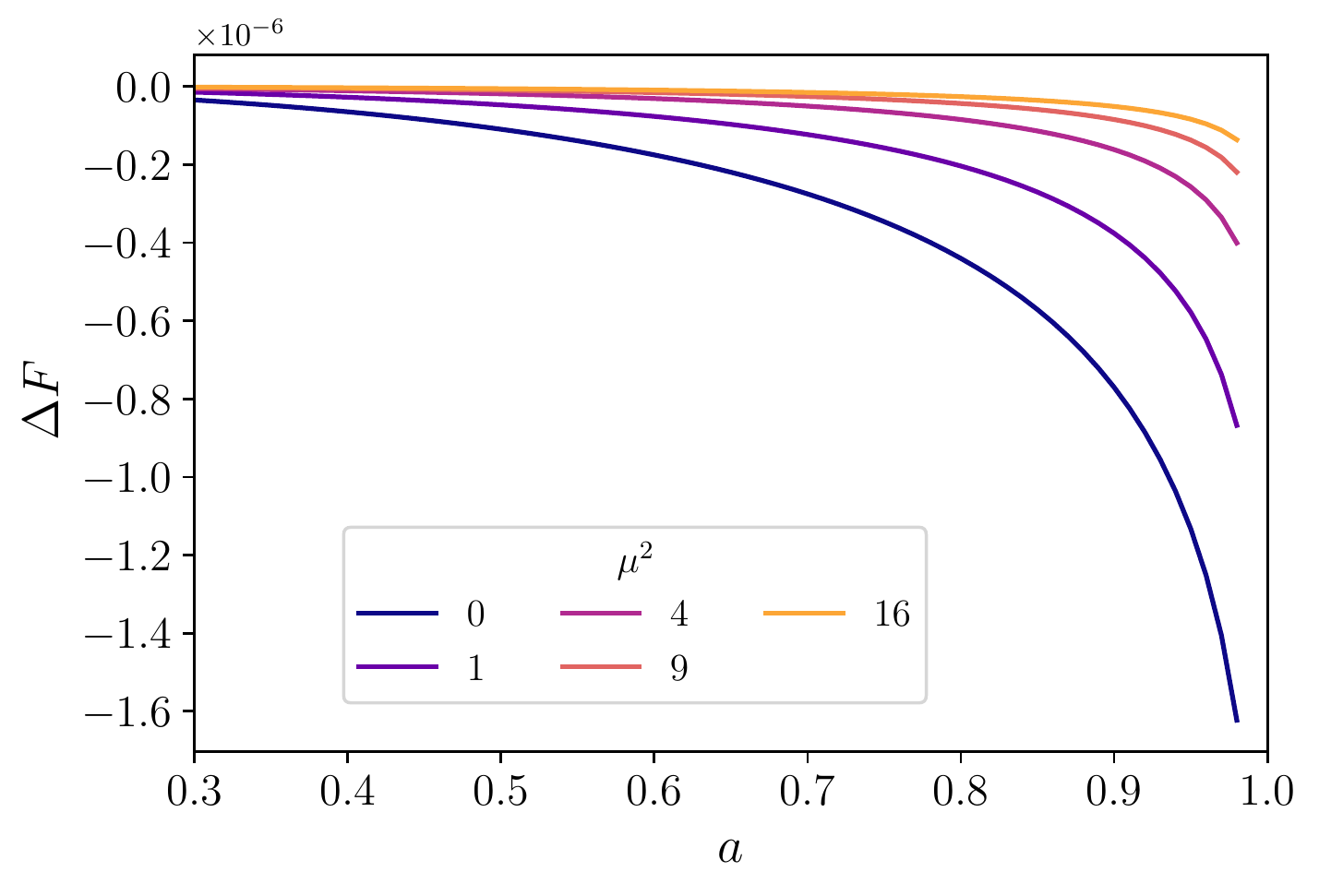}
\caption{Free energy shift vs. black hole angular momentum in Kerr-Newman background. The presented function shows slight drop of free energy for moderate values of angular momentum. however the lowest 
free energy value is found for the extreme black hole regime.}
\label{fig_kn_fa}
\end{figure}

The Kerr-Newman background can be analysed thermodynamically in the similar manner as it has been performed in the Kerr in uniform magnetic field system.
However for the elaborated case,
the free energy dependence on the black hole angular momentum reveals a slightly different behaviour.
These curves are portrayed in Fig. \ref{fig_kn_fa}.
Energy characteristics are monotonic and decreasing with growth of the angular momentum.
Nevertheless in the extreme BH regime the dynamics of the free energy rises and the curves are steeper as $a \rightarrow 1$.
Once again, the curve corresponding to the zero mass has the lowest free energy.

%Energy characteristics possess local minima for moderate values of the black hole angular momentum. 
%Nevertheless the global minimum 
%still lies in the extremal black hole regime, just like in the previously studied background.

%%%%%%%%%%%%%%%%%%%%%%%%%%%%%%%%%%%%%%%%%%%%%%%%%%%%%%%%%%%%%%%%%%%%%%%%%%%%%%%%%
\section{Conclusion}
In our paper we have elaborated the axionic-like  {\it dark matter} model, where the scalar field (axion) are non-trivially coupled to the electromagnetic $U(1)$-gauge
field, via coupling to $\ast F_{\alpha \beta} F^{\alpha \beta}$ invariant. We considered the possibility of accumulating axion {\it dark matter}
clouds in the vicinity of a rotating black hole. Namely we have studied the Kerr black hole immersed in a uniform magnetic field and Kerr-Newman black hole spacetime.
In both cases axion {\it dark matter} clouds tend to accumulate in polar regions of the black hole in question.

As far as Kerr black hole is concerned, it turns out that the increase of a black hole angular momentum does not change the distribution of ultra-light {\it dark matter}
clouds but influences its magnitude. For the increase of the axion mass, the cloud gathers in equatorial area of the object.

For the Kerr-Newman black hole the ultra-light axion {\it dark matter} cloud distribution depends on the black hole angular momentum. The increase of its
value spreads the cloud over the space surrounding the Kerr-Newman black hole.
In the case of large mass axion, the field is strongly dragged towards equatorial area. Moreover, it was revealed that axion-like
{\it dark matter} clouds are preferable for very small, almost zero mass axion fields.

Considered axionic {\it dark matter} clouds do not emerge spontaneously, but are rather magnetically induced. This mechanism naturally requires a magnetic 
component such as galactic magnetic field or a charged rotating black hole. Nevertheless if such {\it dark matter} clouds constitute reality, it will be a 
complicated observational challenge to reveal their existence.
This task will require advanced numerical relativity simulations, which should take into account additional astrophysical and particle-related scenarios, such as e.g.,
plasma-axion {\it dark matter} interactions. We hope to return to
these problems elsewhere.

%\BK{However there is a third case, which somehow merges features of both discussed scenarios.
%Astrophysical black holes, strong radio-sources especially,  usually have an accretion disc.
%Plasma revolving around a BH creates poloidal magnetic field, which can be approximated by uniform magnetic field in the vicinity of BH event horizon.
%Such systems meet requirements of our approach of non-zero $\ast F_{\alpha \beta} F^{\alpha \beta}$ invariant and could be hypothetical habitats for axion-like clouds.}

\appendix
\section{Technical details on the numerical method}

The implemented numerical method relies on the Chebyshev differentiation matrices, which are used to discretize the PDE on a Chebyshev grid.
Similarly to finite difference methods (FDMs), the usage of differentiation matrix allows one to translate a differential equation into a system of linear equations $L \psi = B$.
Unlike the FDM, spectral differentiation requires only few grid points in order to achieve high accuracy.
Having our discrete differential operator constructed, we impose the boundary conditions by substituting particular rows in the matrix and solve the system with standard linear algebra tools. In result we obtain the vector of $\psi$ values on the grid points.
We have implemented the numerical scheme in Python, based on the MATLAB counterpart \cite{matlabnum},  using open source libraries. 

\begin{figure}[h]
\centering
\includegraphics[width=0.45 \textwidth]{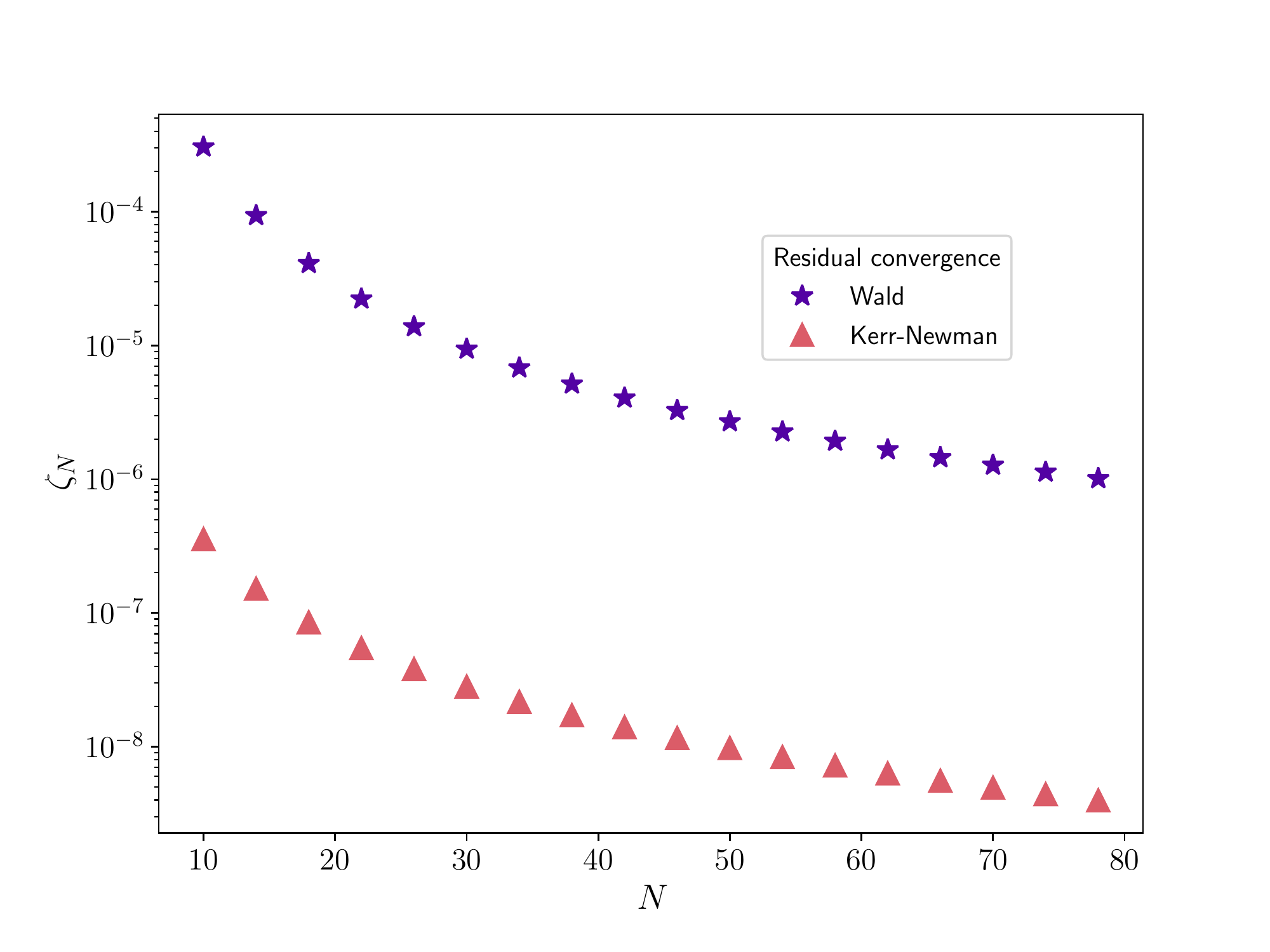}
\caption{Convergence of the mean value of residuals calculated at the set of random points. The error smoothly decays with the growth of the grid.}
\label{fig:conv_res}
\end{figure}

The numerical code has undergone two convergence trials on $N \times N$ grid.  First relies on evaluating the mean of the residuals 
\begin{equation}
\zeta = \langle |L \psi - B| \rangle,
\end{equation}
on a set of random points, which do not belong to the spectral grid.
However we calculate this metric using an another differentiation scheme, in this case standard central finite difference derivative. This allows us to verify if the spectral solutions are relevant \cite{boy00}. The result of this test is presented in Fig.  \ref{fig:conv_res}.
The increase of the number of grid points lowers the error of the solution in both gravitational backgrounds.

\begin{figure}[h]
\includegraphics[width=0.45 \textwidth]{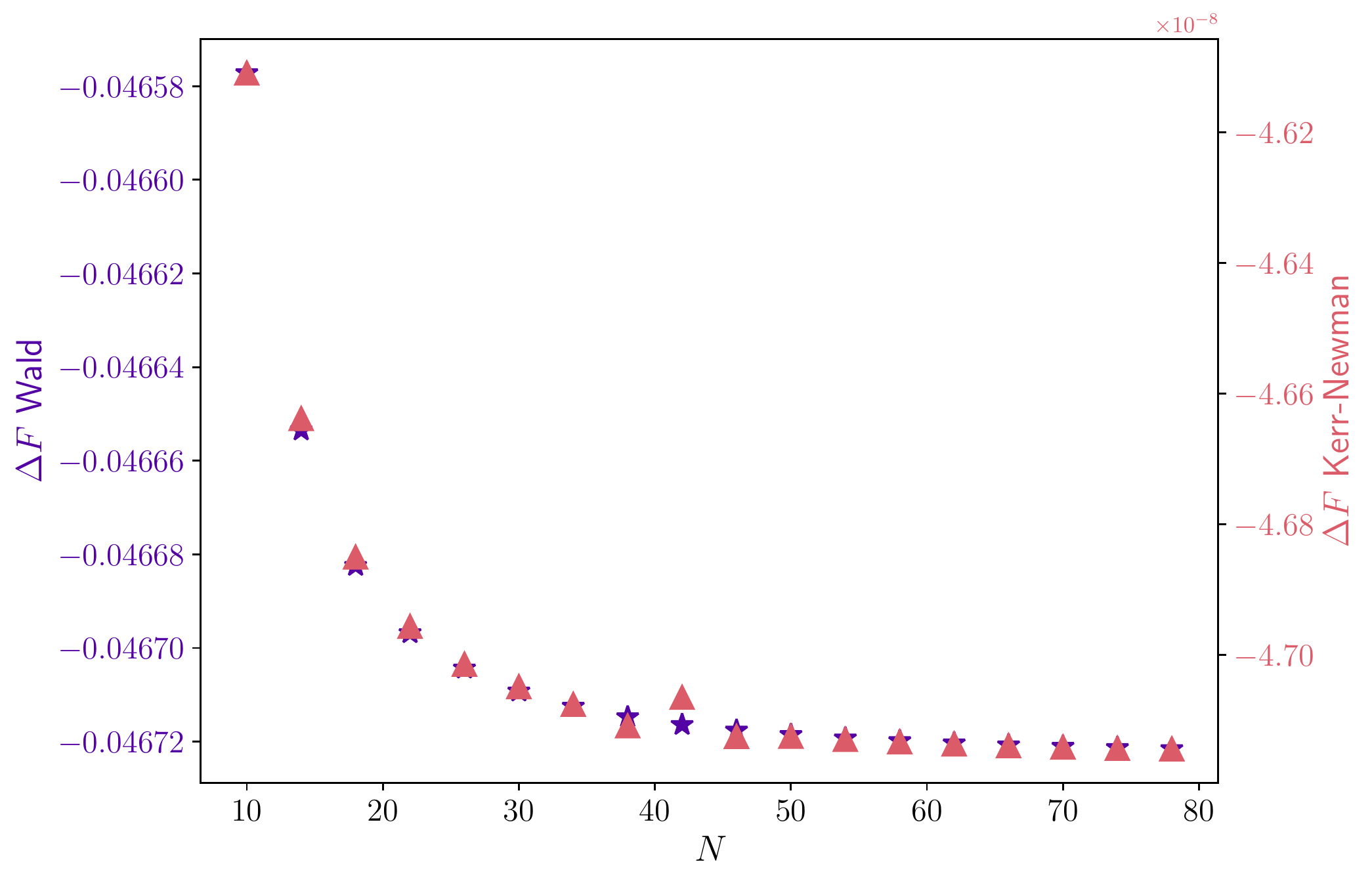}
\caption{Convergence of the free energy of the system.  Left y-scale corresponds to the Wald background (purple stars), while right y-scale to Kerr-Newman solution (red triangles). The value of the free energy quickly converges to the limit.}
\label{fig:conv_f}
\end{figure}

Second numerical test uses the free energy from equation \eqref{eq_freeenergy}.
For the same set of physical parameters we evaluate the free energy, increasing the number of grid points in each step. This test is visualised in Fig.  \ref{fig:conv_f}.
One can see that the free energy quickly converges to its limit.
For convenience we refer to the Kerr black hole in uniform magnetic field as Wald solution.

By the analysis of the convergence of the algorithm we picked $N = 50$ in each direction, as it constitutes a reasonable compromise between the accuracy and the length of computation.  Both presented tests were executed for $\mu = 1$, $a = 0.5$ and $Q = 0.1$ in case of Kerr-Newman background. This is one of the considered cases in Figs. \ref{fig_kw_fa} and \ref{fig_kn_fa}.
For different physical parameters the numerical scheme revealed similar behaviours.
All solutions shown in the plots in this work meet the requirement $\zeta < 10^{-5}$.

\begin{figure}[h]
\includegraphics[width=0.45 \textwidth]{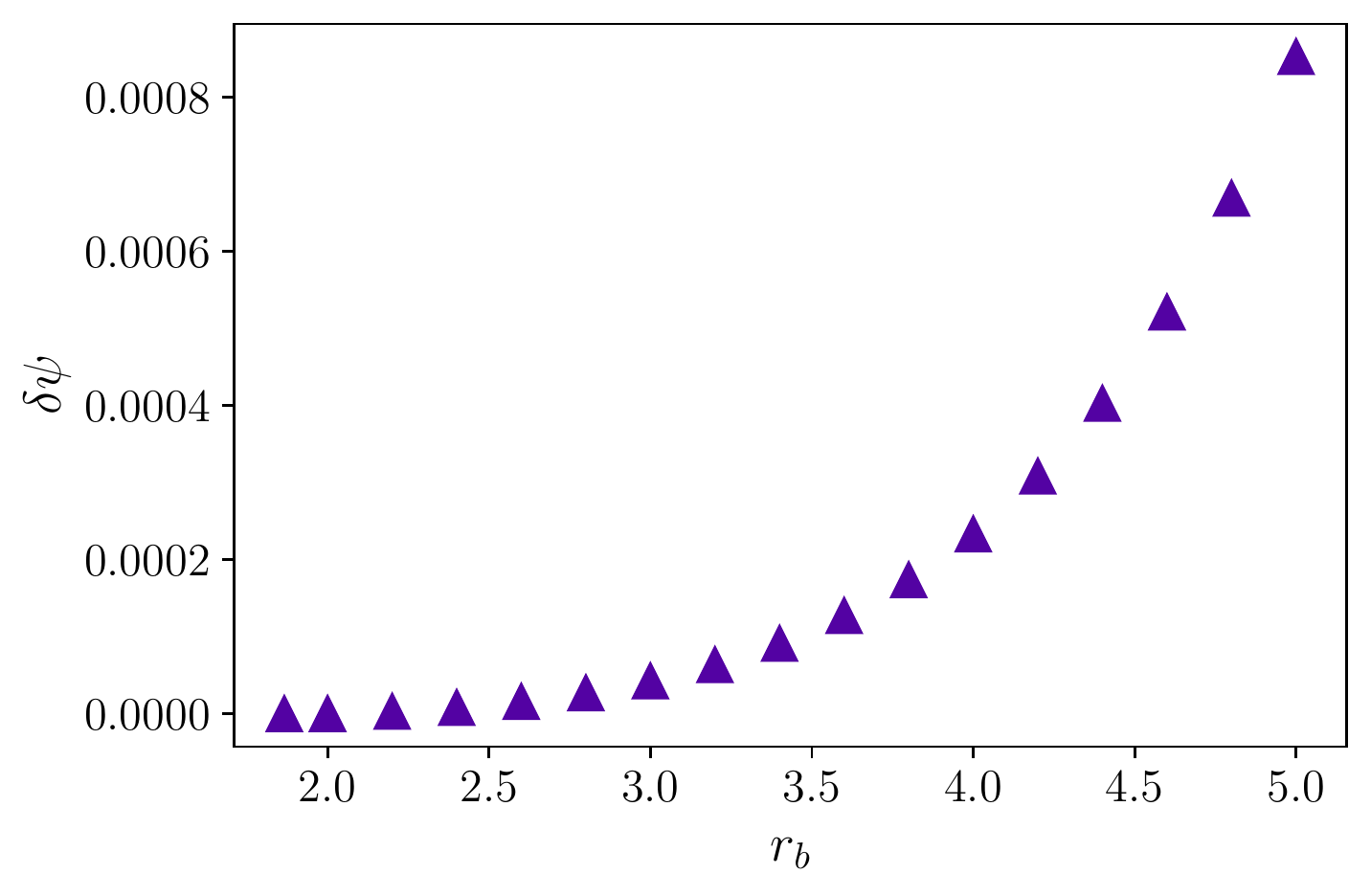}
\caption{Dependence of the value of solution at a distant point in the numerical domain on the location of the inner boundary. The event horizon radius is $r_+ \approx 1.86$ for this case.}
\label{fig:conv_ib}
\end{figure}

Finally we check if the numerical solution is invariant for any change of the location of the inner boundary.
In the whole work we always use the black hole event horizon as the inner boundary $r_b = r_+$ of our numerical domain.
However to make sure that our results are reliable, one can shift the inner boundary arbitrarily and check the behaviour of the solution in the deep interior of the numerical domain.
This is being tested by calculating the value of $\psi$ at the point $\eta(r = 10, ~\theta = \pi/4)$, systematically for subsequent inner boundary locations.
We define a ratio
\begin{equation}
\delta \psi = \frac{\psi(\eta)_{r_b} - \psi(\eta)_{r_b = r_+}}{\psi(\eta)_{r_b = r_+}},
\end{equation}
and compute it for several values of the position of inner boundary $r_b$.
The result of this test is presented in Fig. \ref{fig:conv_ib}.
The value of the function very weakly depends on the position of the inner boundary.
Relative error is to the order of $10^{-4}$ within huge changes of $r_b$ -- up to $3r_+$.

%%%%%%%%%%%%%%%%%%%%%%%%%%%%%%%%%%%%%%%%%%%%%%%%%%%%%%%%%%%%%%%%%%%%%%%%%%%%%%%

\begin{thebibliography}{99}

%%%%%%%%%%%%%%%%%%%%%
%
\def\cmp#1#2#3#4{\emph{#4}, \emph{ Commun. Math. Phys.} {\bf #1} (#3) #2}
\def\lmp#1#2#3#4{\emph{#4}, \emph{ Lett. Math. Phys.} {\bf #1} (#3) #2}
\def\hpa#1#2#3#4{\emph{#4}, \emph{ Hell. Phys. Acta} {\bf #1} (#3) #2}
\def\grg#1#2#3#4{\emph{#4}, \emph{ Gen. Rel. Grav.} {\bf #1} (#3) #2}
\def\pr#1#2#3#4{\emph{#4}, \emph{ Phys. Rev.} {\bf #1} (#3) #2}
\def\prl#1#2#3#4{\emph{#4}, \emph{ Phys. Rev. Lett.} {\bf #1} (#3) #2}
\def\prd#1#2#3#4{\emph{#4}, \emph{ Phys. Rev. D} {\bf #1} (#3) #2}
\def\prb#1#2#3#4{\emph{#4}, \emph{ Phys. Rev. B} {\bf #1} (#3) #2}
\def\prx#1#2#3#4{\emph{#4}, \emph{ Phys. Rev. X} {\bf #1} (#3) #2}
\def\pl#1#2#3#4{\emph{#4}, \emph{ Phys. Lett.} {\bf #1} (#3) #2}
\def\pla#1#2#3#4{\emph{#4}, \emph{ Phys. Lett. A} {\bf #1} (#3) #2}
\def\plb#1#2#3#4{\emph{#4}, \emph{ Phys. Lett. B} {\bf #1} (#3) #2}
\def\prep#1#2#3#4{\emph{#4}, \emph{ Phys. Reports} {\bf #1} (#3) #2}
\def\phys#1#2#3#4{\emph{#4}, \emph{ Physica} {\bf #1} (#3) #2}
\def\jcp#1#2#3#4{\emph{#4}, \emph{ J. Comput. Phys.} {\bf #1} (#3) #2}
\def\jmp#1#2#3#4{\emph{#4}, \emph{ J. Math. Phys.} {\bf #1} (#3) #2}
\def\jpm#1#2#3#4{\emph{#4}, \emph{ J. Phys. A: Math. Gen.} {\bf #1} (#3) #2}
\def\cpr#1#2#3#4{\emph{#4}, \emph{ Computer Phys. Rept.} {\bf #1} (#3) #2}
\def\cqg#1#2#3#4{\emph{#4}, \emph{ Class. Quant. Grav.} {\bf #1} (#3) #2}
\def\cma#1#2#3#4{\emph{#4}, \emph{ Computers Math. Applic.} {\bf #1} (#3) #2}
\def\mc#1#2#3#4{\emph{#4}, \emph{ Math. Compt.} {\bf #1} (#3) #2}
\def\apj#1#2#3#4{\emph{#4}, \emph{ Astrophys. J.} {\bf #1} (#3) #2}
\def\apjs#1#2#3#4{\emph{#4}, \emph{ Astrophys. J. Suppl.} {\bf #1} (#3) #2}
\def\apjl#1#2#3#4{\emph{#4}, \emph{ Astrophys. J. Lett.} {\bf #1} (#3) #2}
\def\acta#1#2#3#4{\emph{#4}, \emph{ Acta Astronomica} {\bf #1} (#3) #2}
%%%%%%%%%%%%%%%%%%%%%%%%%%%%%%%%%%%%%%%%%%%%%%%%%%%%%%%%%%%%%%%%%%%%%%%%%%
\def\apl#1#2#3#4{\emph{#4}, \emph{ Ann. Physik. (Leipzig)} {\bf #1} (#3) #2}
\def\amjp#1#2#3#4{\emph{#4}, \emph{Am. J. Phys.} {\bf #1} (#3) #2}
\def\anp#1#2#3#4{\emph{#4}, \emph{ Ann. Phys.} {\bf #1} (#3) #2}
\def\sa#1#2#3#4{\emph{#4}, \emph{ Sov. Astro.} {\bf #1} (#3) #2}
\def\sia#1#2#3#4{\emph{#4}, \emph{ SIAM J. Sci. Statist. Comput.} {\bf #1} (#3) #2}
\def\aa#1#2#3#4{\emph{#4}, \emph{ Astron. Astrophys.} {\bf #1} (#3) #2}
\def\mnras#1#2#3#4{\emph{#4}, \emph{ Mon. Not. R. Astr. Soc.} {\bf #1} (#3) #2}
\def\npb#1#2#3#4{\emph{#4}, \emph{ Nucl. Phys. B} {\bf #1} (#3) #2}
\def\npa#1#2#3#4{\emph{#4}, \emph{ Nucl. Phys. A} {\bf #1} (#3) #2}

\def\prsla#1#2#3#4{\emph{#4}, \emph{ Proc. R. Soc. London, Ser. A} {\bf #1} (#3) #2}
\def\jhep#1#2#3#4{\emph{#4}, \emph{ JHEP} {\bf #1} (#2) #3}
\def\jcap#1#2#3#4{\emph{#4}, \emph{ JCAP} {\bf #1} (#2) #3}

\def\nuca#1#2#3#4{\emph{#4}, \emph{ Nuovo Cimento A } {\bf #1} (#3) #2}
\def\nucb#1#2#3#4{\emph{#4}, \emph{ Nuovo Cimento B } {\bf #1} (#3) #2}
\def\ijmp#1#2#3#4{\emph{#4}, \emph{ Int. J. Mod. Phys. D} {\bf #1} (#3) #2}
\def\atmp#1#2#3#4{\emph{#4}, \emph{ Adv. Theor. Math. Phys.} {\bf #1} (#3) #2}
\def\ptps#1#2#3#4{\emph{#4}, \emph{ Prog. Theor. Phys. Suppl.} {\bf #1} (#3) #2}
\def\ptp#1#2#3#4{\emph{#4}, \emph{ Prog. Theor. Phys.} {\bf #1} (#3) #2}
\def\lmp#1#2#3#4{\emph{#4}, \emph{ Lett. Math. Phys.} {\bf #1} (#3) #2}
\def\cpam#1#2#3#4{\emph{#4}, \emph{ Comm. Pure Appl. Math.}  {\bf #1} (#3) #2}
\def\adv#1#2#3#4{\emph{#4}, \emph{ Adv. Phys.}  {\bf #1} (#3) #2}
\def\zh#1#2#3#4{\emph{#4}, \emph{ Zh. Eksp. Teor. Fiz.}  {\bf #1} (#3) #2}
\def\mplb#1#2#3#4{\emph{#4}, \emph{ Mod. Phys. Lett. B} {\bf #1} (#3) #2}
\def\jams#1#2#3#4{\emph{#4}, \emph{ J. Austral. Math. Soc. B} {\bf #1} (#3) #2}
\def\appa#1#2#3#4{\emph{#4}, \emph{ Acta Phys. Polonica A} {\bf #1} (#3) #2}
\def\appb#1#2#3#4{\emph{#4}, \emph{ Acta Phys. Polonica B} {\bf #1} (#3) #2}
\def\nat#1#2#3#4{\emph{#4}, \emph{Nature} {\bf #1} (#3) #2}
\def\natcom#1#2#3#4{\emph{#4}, \emph{Nature Commun.} {\bf #1} (#3) #2}
\def\natphys#1#2#3#4{\emph{#4}, \emph{Nature Physics} {\bf #1} (#3) #2}
\def\natmat#1#2#3#4{\emph{#4}, \emph{Nature Mat.} {\bf #1} (#3) #2}


\def\science#1#2#3#4{\emph{#4}, \emph{Science} {\bf #1} (#3) #2}
\def\sciadv#1#2#3#4{\emph{#4}, \emph{Sci. Adv.} {\bf #1} (#3) #2}

\def\arcmp#1#2#3#4{\emph{#4}, \emph{Annual Rev. of Cond. Matter Physics} {\bf #1} (#3) #2}
\def\zphys#1#2#3#4{\emph{#4}, \emph{Z. Phys.} {\bf #1}, (#3) #2}
\def\ncs#1#2#3#4{\emph{#4}, \emph{Nuovo Cimento Suppl.} {\bf #1} (#3) #2}
\def\physb#1#2#3#4{\emph{#4}, \emph{Physica B} {\bf #1}, (#3) #2}
\def\jpcm#1#2#3#4{\emph{#4}, \emph{J. Phys.: Condens. Matter } {\bf #1} (#3) #2}
\def\pnas#1#2#3#4{\emph{#4}, \emph{Proc. Nat. Academy Sciences} {\bf #1} (#3) #2}
\def\sssr#1#2#3#4{\emph{#4}, \emph{Izv. Akad Nauk SSSR, ser. fiz.} {\bf #1} (#3) #2}
\def\jpg#1#2#3#4{\emph{#4}, \emph{ J. Phys. G} {\bf #1} (#3) #2}
\def\chinpb#1#2#3#4{\emph{#4}, \emph{Chin. Phys. B} {\bf #1} (#3) #2}
\def\njp#1#2#3#4{\emph{#4}, \emph{ New J. Phys.} {\bf #1} (#3) #2}
\def\frontphys#1#2#3#4{\emph{#4}, \emph{ Front. Phys.} {\bf #1} (#3) #2}
\def\epl#1#2#3#4{\emph{#4}, \emph{ EPL} {\bf #1} (#3) #2}
\def\rmp#1#2#3#4{\emph{#4}, \emph{ Rev. Mod. Phys.} {\bf #1} (#3) #2}


%Izv. Akad Nauk SSSR, ser. fiz. 
\def\hepph#1#2{{ hep-ph }{#1} (#2)}
\def\arxiv#1#2#3{\emph{#3},{ arXiv }{#1} (#2)}
\def\hepth#1#2{{ hep-th }{#1} (#2)}
\def\grqc#1#2{{ gr-qc }{#1} (#2)}
\def\ibid#1#2#3#4{\emph{#4}, {\it ibid.} {\bf #1} (#3) #2}
\def\conphy#1#2#3#4{\emph{#4}, \emph{Contemporary Physics} {\bf #1}, (#3) #2}
\def\ppnp#1#2#3#4{\emph{#4}, \emph{ Prog. Part. Nucl. Phys} {\bf #1} (#3) #2}
\def\arnps#1#2#3#4{\emph{#4}, \emph{ Annu. Rev. Nucl. Part. Sci.} {\bf #1} (#3) #2}
%%%%%%%%%%%%%%%%%%%%%%%%%%%%%%%%%%%%%%%%%%%%%%%%%%%%%%%%%%%%%%%%%%%%%
%%%%%%%%%%%%%%%%%%%%%%%%CP AXION %%%%%%%%%%%%%%%%%%%%%%%%%%%%%%%%%%%%%%%%%%%%%%%%%%%%%%%%%%%%%%%%%%
\bibitem{pec77}
R. D. Peccei and H. R. Quinn, \prl{38}{1440}{1977}{CP Conservation in the Presence of Pseudoparticles}.
\bibitem{wei78}
S. Weinberg, \prl{40}{223}{1978}{A New Light Boson?}.
\bibitem{wil78}
F. Wilczek, \prl{40}{279}{1978}{Problem of Strong $P$ and $T$ Invariance in the Presence of Instantons}.
%%%%%%%%%%%STRING AXION %%%%%%%%%%%
\bibitem{svr06}
P. Svrcek and E. Witten, \jhep{06}{2006}{051}{Axions in string theory}.
%%%%%%%%%%%%%



\bibitem{fed19}
M. A. Fedderke, P. W. Graham, and S. Rajendram, \prd{100}{015040}{2019}{Axion dark matter detection with CMB polarization}.
\bibitem{co19}
R. T. Co, A. Pierce, Z. Zhang, and Y. Zhao, \prd{99}{075002}{2019}{Dark photon dark matter produced by axion oscillations}.
\bibitem{hua18}
F. P. Huang, K. Kadota, T. Sekiguchi, and H. Tashiro, \prd{97}{123001}{2018}{Radio telescope search for the resonant conversion of cold dark matter axions from 
magnetized astrophysical sources}.
\bibitem{fos20}
J. W. Foster et al., \prl{125}{171301}{2020}{Green Bank and Effelsberg radio telescope searches for axion dark matter conversion in neutron star magnetospheres}.
\bibitem{sen18}
S. Sen, \prd{98}{103012}{2018}{Plasma effects on lasing of a uniform ultralight axion condensate}.
\bibitem{ros18}
J. G. Rosa and T. W. Kephart, \prl{120}{231102}{2018}{Stimulated axion decay in superradiant clouds around primordial black holes}.
\bibitem{gar18}
B. Garbrecht and J. I. Mc Donald, \jcap{07}{2018}{044}{Axion configurations around pulsars}.

\bibitem{dar20a}
J. Darling, \apjl{900}{L28}{2020}{New limits on axionic dark matter using the magnetar PSR J1745-2900}.
\bibitem{dar20b}
J. Darling, \prl{125}{121103}{2020}{Search for axion dark matter using the magnetar PSR J1745-2900}.
\bibitem{ari12}
P. Arias, D. Cadamuro, M. Goodsell, J. Jaeckel, J. Redondo, and A. Ringwald, \jcap{06}{2012}{013}{WISPy cold dark matter}.
\bibitem{gra15}
P. W. Graham, I. G. Irastorza, S. K. Lamoreaux, A. Lindner, and K. A. van Bibber, \arnps{65}{485}{2015}{Experimental searches for axion and axion-like particles}.





\bibitem{pla18}
A. D. Plascencia and A. Urbano, \jcap{04}{2018}{059}{Black hole supperadiance and polarization-dependent bending of light}.

\bibitem{gao20}
X. Bi, Y. Gao, J. Guo, N. Houston, T. Li, F. Xu, and X. Zhang, \arxiv{2002.01796}{2020}{Axion and dark photon limits from Crab Nebula high energy gamma-rays}.

\bibitem{ike19}
T. Ikeda, R. Brito, and V. Cardoso, \prl{122}{081101}{2019}{Blasts of light from axions}.
\bibitem{bos19}
M. Boskovic, R. Brito, V. Cardoso, T. Ikeda, and H. Witek, \prd{99}{035006}{2019}{Axionic instabilities and new black hole solutions}.
\bibitem{car18}
V. Cardoso, O. J. C. Dias, G. S. Harnett, M. Middleton, P. Pani, and J. E. Santos, \jcap{03}{2018}{043}{Constraining the mass of dark photons and axion-like particles through black-hole superradiance}.


\bibitem{wal74}
R. M. Wald, \prd{10}{1680}{1974}{Black hole in a uniform magnetic field}.






\bibitem{kic20}
B. Kiczek and M. Rogatko, \prd{101}{084035}{2020}{Influence of dark matter on black hole scalar hair}.

\bibitem{matlabnum} Lloyd N. Trefethen, \emph{Spectral methods in MATLAB}, SIAM, Philadelphia, (2000).


\bibitem{gar19}
A. Gracon {\it et al.}, \sciadv{5}{eaax4539}{2019}{Constraints on bosonic dark matter for ultralow-field nuclear magnetic resonance}.
\bibitem{una20}
C. \"Unal, F. Pacucci, and A. Loeb, \arxiv{2012.12790}{2020}{Properties of ultralight bosons from heavy quasar spins via superradiance}.

\bibitem{boy00}
J. P. Boyd, \textit{Chebyshev and Fourier Spectral Methods}, DOVER Publications Inc, Mineola, New York (2000).









\end{thebibliography}
\end{document}